\documentclass[12pt, a4paper]{article}
\usepackage[square,sort,comma,numbers]{natbib}

\usepackage[latin1]{inputenc}
\usepackage{amsmath,amsfonts, amsthm, amssymb}
\usepackage{newlfont}
\usepackage{fancyhdr}
\usepackage[Glenn]{fncychap}
\usepackage{fullpage}

\usepackage[T1]{fontenc}
\usepackage{array,multicol}
\usepackage[pdftex]{graphicx}
\usepackage{framed}
\usepackage{setspace}
\usepackage{listings}
\usepackage{tikz}
\usepackage{pgfplots}
\usepackage{tikz-3dplot}
\usepackage{booktabs}
\usepackage{bm}
\usepackage{siunitx}

\tdplotsetmaincoords{70}{150}
\pgfplotsset{compat=newest}

\usepackage{setspace} % Espaciado entre lineas
\def\phi{\varphi}
\def\RR{\mathbb R}

\def\incep{\left\{\begin{array}{ll} }
 \def\termin{\end{array}\right. }

\def\la2{\lambda^2}

\newcommand{\bss}{{\boldsymbol s}}

\newcommand{\btt}{{\boldsymbol t}}
\newcommand{\bnu}{{\boldsymbol \nu}}

\newcommand{\bmm}{{\boldsymbol m}}
\newcommand{\br}{{\boldsymbol r}}
\newcommand{\bR}{{\boldsymbol R}}
\newcommand{\be}{{\boldsymbol e}}
\newcommand{\bc}{{\boldsymbol c}}
\newcommand{\bv}{{\boldsymbol v}}

\newcommand{\bF}{{\boldsymbol F}}
\newcommand{\bK}{{\boldsymbol K}}

\newcommand{\nn}{{\boldsymbol e}_3}

\newcommand{\bb}{{\boldsymbol b}}
\newcommand{\bx}{{\boldsymbol x}}
\newcommand{\U}{{\cal U}}
\newcommand{\T}{{\cal T}}

\newcommand{\K}{{\cal K}}

\newcommand{\bN}{\boldsymbol  N}

\newcommand{\bM}{\boldsymbol  M}
\newcommand{\bE}{{\boldsymbol  E}}

\newcommand{\bU}{\boldsymbol {\cal U}}
\newcommand{\bCC}{\boldsymbol {\cal C}}
\newcommand{\bMM}{\boldsymbol {\cal M}}
\newcommand{\bAA}{{\boldsymbol A}}

\newcommand{\bni}{{\boldsymbol n}}
\newcommand{\bI}{{\boldsymbol I}}
\newcommand{\bS}{{\boldsymbol S}}

\newcommand{\Oo}{ {\cal O}}

\def\R{\mathbb{R}}

\newcommand{\bT}{{\boldsymbol {\cal T}}}
\newcommand{\bTT}{{\boldsymbol T}}

\newcommand{\bSS}{{\boldsymbol{\cal S}}}
\newcommand\bC{\boldsymbol{C}}

\newcommand\brX{\bar{X}}

\numberwithin{equation}{section}

\usepackage{marginnote}

\usepackage{hyperref}
\hypersetup{
  colorlinks   = true,    % Colours links instead of ugly boxes
  urlcolor     = blue,    % Colour for external hyperlinks
  linkcolor    = blue,    % Colour of internal links
  citecolor    = blue     % Colour of citations
}
\begin{document}

\title{Design of pre-stressed plate-strips to cover non-developable shells} 

\author{Alexandre Danescu\footnote{University of Lyon, Institute of Nanotechnologies - INL, UMR CNRS 5270, Ecole Centrale de Lyon, Ecully, France, \href{mailto:alexandre.danescu@ec-lyon.fr}{\nolinkurl{alexandre.danescu@ec-lyon.fr}}} \ and Ioan R. Ionescu\footnote{LSPM,   University Sorbonne-Paris-Nord, Villetaneuse, France,  and IMAR, Romanian Academy, Bucharest, Romania, \href{mailto:ioan.r.ionescu@gmail.com}{\nolinkurl{ioan.r.ionescu@gmail.com }}}}
\date{version : \today}
\maketitle

\begin{abstract}

In this paper we address the following design problem: what is the shape of a plate and the associated pre-stress that relaxes toward a given three-dimensional shell ? As isometric transformations conserve the gaussian curvature, three-dimensional non-developable shells cannot be obtained from the relaxation of pre-strained plates by using isometric transformations only. Overcoming this geometric restriction, including small-strains and large rotations, solves the problem for small areas only. This paper dispenses with the small-area restriction to cover three-dimensional shells fully by using shell-strips. Since shell-strips have an additional geometric parameter, we show that under suitable assumptions that relate the width of the strip to the curvature of the shell, we are able to design arbitrary shell surfaces by covering them with shell-strips. As an illustration,
we provide optimized covers of the sphere in a variety of different surface-strips relaxed from plate-strips with homogeneous and isotropic pre-stress. 
Moveover, we propose  the design of the torus, of the helicoid and of the non-developable M\"obius band, which requires inhomogeneous and anisotropic pre-stress. 

\end{abstract}

{\bf  Keywords}: nonlinear elasticity, large rotations and small strains, shell design, pre-stress, non-developable strip-surfaces.
\newpage

\tableofcontents{}
\bigskip

%{\bf  Highlights}

%\begin{enumerate}
	
%	\item nnn
	
%	\end{enumerate}
%\newpage

\section{Introduction}

Modern technological developments for semiconductors at the nanoscale, such as Molecular Beam Epitaxy (MBE), allow very fine tuning of the material properties through control of the lattice parameter, while the geometry of the domain can be easily implemented by using photolithography. However, the lattice mismatch between the planar template and the grown crystal should be kept as small as possible and the fabrication technology at the nano-scale is essentially planar. But from a practical point of view, the bending process of a pre-stressed multi-layer material may be beneficial, as one can use it to design various three-dimensional objects starting from planar pre-strained templates, thus encompassing the planar technology.

Since the semiconductors are anisotropic elastic/brittle materials, the study of the equilibrium shape for a bilayer plate of a given geometry can be formulated as a classical linear or nonlinear direct elasticity problem. Given either a specific or a generic elastic energy, one looks for the existence/uniqueness/approximation of configurations that realizes the local (or global) minimum of the considered elastic energy. For the modeling aspects in thin layer pre-stressed materials, asymptotic models inspired by dimension reduction were proposed in \cite{fox1993, dret_raoult_95, friesecke2002theorem, friesecke2002, friesecke2006, de2020energy, wang2019uniformly, de2019hierarchy}. The linear approach cannot recover experimental results illustrated, for instance, in \cite{prinz2017} for the relaxed configurations that do not fit the small perturbation theory, which is a situation frequently encountered in applications. This is a benchmark for various asymptotic and/or exact nonlinear models for elastic plates/shells, and recent results \cite{lewicka2017,lewicka2018} on various approximations of the three-dimensional elasticity with incompatible pre-strain/stress provide a hierarchy of non-linear elastic models.

However, the classical approach cannot answer the following important practical question: {\em what is the two-dimensional shape that relaxes toward a given three-dimensional surface/object?}  For problems of design by stress relief for a given target geometry, we are looking for an elastic, pre-stressed material and a reference geometry so that the target geometry represents the naturally relaxed configuration of the reference geometry.
The minimal mechanical and geometrical setting needed to address  the general design problem mentioned above is that of small strains but large rotations. The motivation of the small strains assumption relies mainly on the fact that the single source of elastic energy is the small pre-strain/pre-stress,  while the large rotations framework is needed, since in most situations the planar design contains a large characteristic length. Moreover, since we are focusing on brittle-elastic materials (such as semiconductors), the small deformations assumption is merely a {\em technological restriction} and not a mathematical simplification.

Previous results \cite{seleznev2003, prinz2003, prinz2006, prinz2001, prinz2000, prinz2017} concerning relaxation of pre-stressed bilayer materials focus on straight ribbons that relax toward rolls and curls, all based on isometric transformations. However, it is well-known that the class of isometries between planar and three-dimensional surfaces, extensively studied in \cite{fosdick2016}, is too narrow to cover simple non-developable surfaces occurring in pre-stressed relaxation design problems. To circumvent this theoretical drawback, in a recent paper \cite{danescu2020shell} we developed a shell design model built on a non-isometrical perturbation assumption (of Love-Kirchhoff type superposed on a plate-to-shell theory \cite{steigmann2013, ciarlet2018, steigmann2007thin, steigmann2007asymptotic, steigmann2014classical}). The geometric description involves a single small parameter $\delta\ll 1$,  the product between the thickness of the shell and its curvature.

The main difficulty in applying a shell design model \cite{danescu2020shell} is of a geometric nature. Indeed, for several common mid-surfaces the small-strain assumption drastically reduces the surface width. For instance, only small parts of a spherical shell can be recovered from the pre-stressed plates as shown in Fig. 4 in \cite{danescu2020shell}.  To encompass this limitation, in this paper we construct another type of shell, called a strip-shell, for which this assumption can be fulfilled by an appropriate choice of an additional geometric parameter, namely the strip width. The geometries of shell-strips introduce an additional small parameter, further denoted $\eta$, which is the ratio between the width of the strip and the curvature radius. Then, for $\delta = \eta^2$ the assumptions of plate-to-shell theory \cite{danescu2020shell} are fulfilled and for any strip of a given shell we obtain a simple model to design the corresponding plate-strip (i.e., to compute the shape and pre-stress moment of the plate). The next step is to cover the given surface (shell) with one or several strips, for which we can design the corresponding planar (plate) strips.

The design problem we address here was also previously considered in \cite{aharoni2014,aharoni2018,griniasty2019,vanress2017} in a different conext where the in-plane average pre-stress related to growth was considered in contrast to our thickness variation approach suitable for planar fabrication technology.

This paper is structured as follows: the second section presents the basic geometric, kinematic and constitutive assumptions and a simplified version of the plate-to-shell model for the design obtained in \cite{danescu2020shell}. The third section addresses the particular problem of the design for a shell-strip from a pre-stressed plate-strip. As previously mentioned, by requiring the ratio between the shell-strip thickness and the shell-strip width to be $\Oo(\sqrt{\delta})$, we can use the shell-to-plate theory to obtain the pre-stress bending moment. More exactly, we define a second-order  strip constructed along an arbitrary curve of the shell mid-surface   and its Lagrangian counterparts, a planar strip along a planar curve. We prove that  the natural mapping  of the  plate-strip to a shell-strip  is a small strain transformation  if the curvature of the planar strip coincides with the  geodesic curvature of the shell-strip  and some other additional conditions on the width are fulfilled. The design problem we address includes much more than the class of isometric transformation so that, under specific assumptions, we are able to design shells for which the Gaussian curvature of the mid-surface does not vanish. The fourth section offers three examples in this class: three designs that completely cover the sphere, two designs that partially cover the torus and, finally,  two examples of rotoidal strips (helicoid and classical M\"obius ribbon).

\section{Plate-to-shell equations for design}

In this section,  we present a simplified version of the
model obtained in \cite{danescu2020shell} for plate-to-shell design. In the applications we have in mind, the pre-stress/strain has an
important thickness heterogeneity but the material properties are almost homogeneous.  Here, this allows us to consider only materials with a weak transversal material heterogeneity. In this case, an important simplification of the model appears: the six equations for the design problem can be decoupled in two families.  Three of them involve the small perturbation which can be computed from the membrane deformation and the pre-stress resultant, and will not be used in the design problem. The remaining three, also called the design equations, involve the three components of the pre-stress moments and the membrane curvature.

\subsection{Geometric assumptions}

Let $\bss_0 \subset \RR^3$ be the design Eulerian mid-surface and let $\nn$ denote the unit normal, and $\K$ the curvature tensor acting from the tangent plane $\T$    into itself. The designed shell is the given by 
\begin{equation}  
	\label{ssE}
	\bss=\{\bx_0+x_3\nn(\bx_0) \; ; \;  \bx_0\in \bss_0, x_3\in (-\frac{h}{2},\frac{h}{2}) \},
\end{equation}
where $h=h(\bx_0)$ is the shell thickenss. Let us also consider that 
\begin{equation}
	\label{SSL} 
	\bSS=\{X=(\bar{X},X_3) ; \bar{X}\in \bSS_0\, X_3\in (-\frac{H(\bar{X})}{2},\frac{H(\bar{X})}{2})\}
\end{equation}
is the Lagrangian plate with mid-surface $\bSS_0 \subset \RR^2$ and thickness $H=H(\bar{X})$ in the Lagrangian configuration, where we use $\bar{X}=(X_1,X_2)$.

In what follows, $\delta \ll 1$ will be a small parameter characterizing the Eulerian and Lagrangian shell thickness through 
\begin{equation}\label{H}
	h\vert \K \vert=\Oo(\delta), \quad \frac{H}{L_c}=\Oo(\delta), \quad \vert \nabla_2 H\vert =\Oo(\delta), 
\end{equation}
where $L_c$ is the characteristic length of the surface and $\nabla_2$ is the gradient with respect to $\bar{X}\in \bSS_0$. 

The {\em main geometric assumption} is that there exists a transformation $\bx_0 : \bSS_0 \to \RR^3$   of  the Lagrangian mid-surface $\bSS_0$ into the designed  Eulerian one $\bss_0$ (i.e., $\bss_0 = \bx_0 (\bSS_0)$) such that the associated deformation of the geometric transformation is small, i.e., 
\begin{equation} \label{AG}
	\vert   \bF^T_0(\brX)\bF_0(\brX)-\bI_2  \vert  =\Oo(\delta), \quad \mbox{for all} \;  \brX \in \bSS_0.
\end{equation}
Here,   $\bF_0=\nabla_2  \bx_0$ is the gradient tensor acting in each point $\bar{X}=(X_1,X_2) \in \bSS_0$,  from  $\RR^2$ into the tangent plane $\T(\bx_0(\bar{X}))$ of the designed surface $\bss_0$ and $\bI_2=\bc_1 \otimes\bc_1+\bc_2 \otimes\bc_2 $ is the identity tensor on $\RR^2$ and $\{\bc_1,\bc_2, \bc_3\}$ is the Cartesian basis in the Lagrangian description. We further denote by $\bK=\bF_0^T\K \bF_0$ the Lagrangian curvature tensor acting from $\RR^2$ into itself. 

\bigskip

The kinematics of the plate deformation involves the classical Love-Kirchhoff assumption, i.e.: {\em the normal to the plate mid-plane remains normal to the designed mid-surface} but in a finite deformation context and thus including large rotations. In addition, the transversal deformation is affine with respect to the plate thickness. Superposed to  the kinematics associated to the exact design which reproduces the target mid-surface, we  consider a small perturbation of Love-Kirchhoff type in order to compensate the small (membrane) deformation of the proposed geometric transformation. As a consequence, the mid-surface of this overall kinematics will be close to the designed mid-surface, and for this reason we called it {\em approximative designed kinematics}. Our main goal is to provide conditions which ensure that an approximative designed configuration can be reached by releasing a suitable pre-strained plate.

\subsection{Constitutive  assumptions}

From the constitutive point of view, applications to semiconductor materials require cubic materials under bi-axial pre-strain. 
We also assume weak transverse material heterogeneity that ensures the decoupling between average  and moment equations. This significant simplification of the proposed model is motivated by the fact that, in most of the applications to semicondutor layers grown by MBE, the pre-strain is induced by the fine-tuning of the layers composition (for instance $\textrm{In}_{1-\alpha}\textrm{Ga}_\alpha\textrm{P}$ for small $\alpha$). With these ingredients, we were able to rely on the theoretical predictions with the experimental evidence \cite{danescu2013, danescu2018} for  thin ($\simeq 200$ nm thick) semiconductor pre-strained multi-layers. 

Here we consider a pre-stressed hyper-elastic material undergoing small strains but large rotations. If $\vert\bE\vert = \Oo(\delta)$ is the strain tensor and $\bS$ is the second Piola-Kirchhoff stress tensor, then the constitutive equation reads 
\begin{equation} \label{Elas}
	\bS= \bCC \bE + \bS^*+\Sigma\Oo(\delta^2),
\end{equation}
where $\bCC$ is the fourth-order tensor of elastic coefficients and $\bS^*=\bS^*(\bar{X},X_3)$ is the pre-stress with 
$\vert\bS^* \vert =\Sigma\Oo(\delta)$ and $\Sigma$ is a characteristic stress.

In what follows, we consider only orthotropic materials with the elastic coefficients (Voigt notation) $C_{ij}=C_{ij}(\bar{X},X_3)$, i.e.,
$$\bCC \bAA= \bCC_2 \bAA_2 
+ A_{33}\bC_3
+ (\bC_3: \bAA_2 +C_{33}A_{33})\bc_3 \otimes\bc_3 
+ 4C_{44}A_{23}(\bc_2 \otimes\bc_3)_S
+ 4C_{55}A_{13}(\bc_1 \otimes\bc_3)_S,
$$
where we have denoted by $ \bAA_2$ the in-plane part of the 3-D tensor  $\bAA$ (i.e., $\bAA_2=A_{11}\bc_1 \otimes\bc_1+A_{12}\bc_1 \otimes\bc_2+A_{21}\bc_2 \otimes\bc_1+A_{22}\bc_2 \otimes\bc_2$), by $\bC_3 = C_{13}\bc_1 \otimes\bc_1 +C_{23}\bc_2 \otimes\bc_2,$ and by 
$\bCC_2 \bAA_2=(C_{11}A_{11}+C_{12}A_{22})\bc_1 \otimes\bc_1
+ (C_{12}A_{11}+C_{22}A_{22})\bc_2 \otimes\bc_2
+ 4 C_{66}A_{12}(\bc_1 \otimes\bc_2)_S $. 
If the material is isotropic, then we obtain the Saint-Venant-Kirchhoff law, i.e., (Voigt notation),
\begin{equation}\label{Isotrop}
	\bCC _2\bAA_2=\lambda trace(\bAA_2)\bI_2+2\mu\bAA_2,  \quad \bC_3=\lambda\bI_2, \quad C_{33}=\lambda+2\mu, \quad C_{44}=C_{55}=\mu,
\end{equation}
where $ \lambda =\lambda(\bar{X},X_3)$ and $\mu=\mu(\bar{X},X_3)$ are the Lam\'e elastic moduli.

We assume further that the material has a weak transversal heterogeneity, i.e.,
\begin{equation} \label{aHom}
	\hat{\bCC}_2=\Sigma\Oo(\delta),   \quad  \breve{\bCC}_2=\frac{1}{12}\bar{\bCC}_2+\Sigma\Oo(\delta), \\
\end{equation}
$$
\hat{\bC}_3=\Sigma\Oo(\delta),   \quad  \breve{\bC}_3=\frac{1}{12}\bar{\bC}_3+\Sigma\Oo(\delta),  \quad  \hat{C}_{33}=\Sigma\Oo(\delta), \quad  \breve{C}_{33}=\frac{1}{12}\bar{C}_{33} +\Sigma\Oo(\delta),$$
where we have denoted   by  $\bar{\bAA}, \hat{\bAA}, \breve{\bAA}$  the thickness average and first and second-order moment of  $\bAA$, i.e.,
$$\displaystyle  \bar{\bAA}=\frac{1}{H} \int_{-H/2}^{H/2} \bAA(X_3) dX_3,\quad \hat{\bAA}=\frac{1}{H^2} \int_{-H/2}^{H/2} X_3\bAA(X_3) dX_3, \quad  \breve{\bAA}=\frac{1}{H^3} \int_{-H/2}^{H/2} X_3^2\bAA(X_3) dX_3.$$
For isotropic materials,  the condition of weak transversal heterogeneity (\ref{aHom}) becomes
\begin{equation} \label{IsoWH}
	\hat{\lambda} =\Sigma\Oo(\delta),\quad  \hat{\mu}=\Sigma\Oo(\delta), \quad  \breve{\lambda} =\frac{1}{12}\bar{\lambda}+\Sigma\Oo(\delta),\quad  \breve{\mu}=\frac{1}{12}\bar{\mu}+\Sigma\Oo(\delta). \end{equation}

\bigskip

Concerning the pre-stress $\bS^*$, we assume that the tangential pre-stresses acting on the surfaces parallel to the mid-surface vanish, i.e.,
\begin{equation} \label{ASt}
	\bS^*=\bS^*_2+ S^*_{33} \bc_3 \otimes\bc_3+\Sigma\Oo(\delta^2),
\end{equation}
where $\bS^*_2$ is the in-plane pre-stress and $S^*_{33}$ is the transversal pre-stress.

\subsection{Moment plate-to-shell equations}

We recall here from \cite{danescu2020shell} the moment equations relating the pre-stress $\bS^*$ and the Lagrangian curvature $\bK$. If 
$$
\bM^*=H^2 \hat{\bS^*_2}- \frac{H^2\hat{S}_{33}^*} {\bar{C}_{33}}\bar{\bC}_3,
$$
denotes the pre-stress couple, then the shell boundary value problem for the moments reads 
\begin{equation}  \label{MoTNh}
	div_2(\frac{H^3}{12}\bMM_2\bK+ \bM^*)=0  \quad \mbox{in}  \; \bSS_0,  \quad
	(\frac{H^3}{12}\bMM_2\bK+ \bM^*)\bar{\bnu}=0 \quad \mbox{on} \;  \partial \bSS_0,
\end{equation}
\begin{equation}
	\label{MonNh}   (\frac{H^3}{12}\bMM_2\bK+ \bM^*):\bK =0  \quad \mbox{in}  \; \bSS_0.
\end{equation}
In the above, 
$\bMM_2$ denotes the in-plane elastic fourth-order tensor
$$
\bMM_2 \bAA_2=\bar{\bCC}_2\bAA_2-\frac{\bar{\bC}_3:\bAA_2 } {\bar{C}_{33}} \bar{\bC}_3
$$
which, in the particular case of isotropic materials, becomes
\begin{equation} \label{MM}
	\bMM_2 \bAA_2=\frac{2\bar{\mu}\bar{\lambda}}{\bar{\lambda}+2\bar{\mu}}trace(\bAA_2)\bI_2
	+2\bar{\mu}\bAA_2. 
\end{equation}

\bigskip

Let us note that the equations and boundary conditions (\ref{MoTNh}-\ref{MonNh})  are satisfied if
\begin{equation} \label{M0}
	\bM^*=-\frac{H^3}{12}\bMM_2\bK \quad \mbox{in}  \quad \bSS_0,
\end{equation}
but this is only a sufficient condition for (\ref{MoTNh})-(\ref{MonNh}) and not a necessary one. In the particular case of isotropic materials, from (\ref{M0}) we obtain a simple formula relating the invariants of pre-stress moment $\bM^*$ to the invariants of Lagrangian curvature tensor $\bK$: 
\begin{equation} \label{Ms} 	
	trace(\bM^*)=-\frac{H^3}{6}(\frac{2\bar{\mu}\bar{\lambda}}{\bar{\lambda}+\bar{\mu}}+\bar{\mu})trace(\bK), \quad 
	\vert (\bM^*)^D\vert =\frac{H^3}{6}\bar{\mu}\vert (\bK)^D\vert,
\end{equation}
where $(\bAA)^D=\bAA-\frac{trace(\bAA)}{2}\bI_2$ denotes the two-dimensional deviatoric part of $\bAA$.

\section{Designing a pre-stressed plate-strip for a given shell-strip}

For several common mid-surfaces the small-strain assumption (\ref{AG}) drastically reduces the surface width with respect to the curvature radius and this is the main limitation in applying the above plate-to-shell model. This is why, in this section, we introduce another type of shell, called a strip-shell, for which assumption (\ref{AG}) can be fulfilled by an appropriate choice of an additional geometric parameter, namely the strip width $2d$. From a geometric point of view, a shell-strip has two length scales linking the mid-surface curvature to the strip width and to the shell thickness $h$.  We prove that the small-strain assumption (\ref{AG}) can be satisfied by designing the planar strip and choosing the ratio $h/d= \Oo(\sqrt{\delta})$. Then, we can use the plate-to-shell model described above to obtain the distribution of pre-stress moment $\bM^*$.  

\subsection{Designing strips with small strain}

The main objective of this section is to design an Eulerian surface-strip $\bss_0$ of a given surface $\bU$ along a given curve $\bc \subset \bU$ and a Lagrangian planar strip  $\bSS_0$ such that the small-strain assumption (\ref{AG}) holds.

\subsubsection{Constructing a second-order strip along a curve on a surface} 
\label{Surface}

Let $\bU \subset \RR^3$  be a surface given by its parametric description $u \to \br_\U(u) \in \RR^3$, where $u=(u_1,u_2)$ are coordinates in some subset $\Omega \subset \RR^2$. We denote $\bb_1, \bb_2$ for the covariant basic vectors, $\bb^1, \bb^2$ for the contravariant basic vectors,  $g_{\alpha\beta}={\bb}_{\alpha}\cdot{\bb}_{\beta}$ for the covariant metric tensor and $g^{\alpha\beta}$ for the components of its inverse, the contravariant metric tensor. Let $\be_3=\bb_1\wedge\bb_2/\sqrt{g_{11}g_{22}}$ be the unit normal vector on $\bU$ and let  $\K=\partial_{u_i}\be_3\otimes\bb^i$ be the curvature tensor.  

\begin{figure}[ht!]
\centering
\begin{tikzpicture}
    \node[anchor=south west,inner sep=0] at (0,0) 
    {\includegraphics[width=16cm]{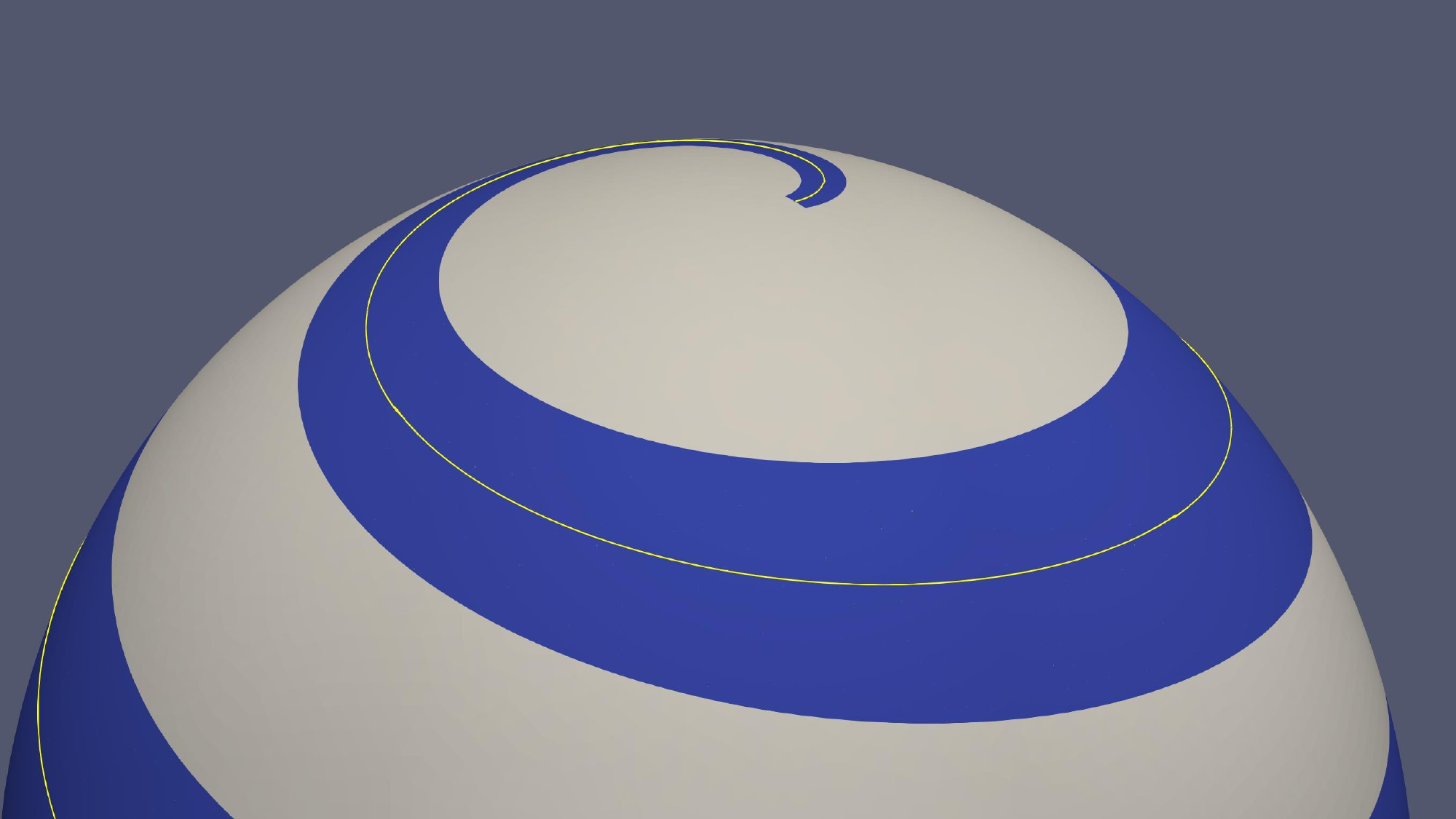}};
    \draw[->,>=stealth,color=yellow,thick] (8,2.7) -- (9,2.5) node[below] {$t_0$};
    \draw[->,>=stealth,color=yellow,thick] (8,2.7) -- (8.7,3.2) node[right] {$n_0$};
    \draw[->,>=stealth,color=yellow,thick] (8,2.7) -- (8,3.5) node[above] {$m_0$};
    \draw[->,>=stealth,color=red,thick] (8,2.7) -- (7.7,3.4) node[left] {$e_3$};
    % strip frame
    \draw[thick,yellow] ([shift=(-10:0cm)]6.5,4.35) arc (132:158:5.43cm);
    \draw[->,>=stealth,color=green,thick] (5.7,3.45) -- (6.7,2.9) node[below] {$b_s$};
    \draw[->,>=stealth,color=green,thick] (5.7,3.45) -- (6.2,4.4) node[left] {$b_q$};
    \draw[->,>=stealth,color=red,thick] (5.7,3.45) -- (5.4,4.2) node[left] {$e_3$};
    \node[right,color=yellow,yshift=-0.3cm] (C) at (11.5,3.5) {$\bCC$};
    \node[right,color=green,yshift=-0.3cm] (D) at (12.7,2.5) {$\bss_0$};
    % surface frame
    \draw[->,>=stealth,color=red,thick] (6.7,6.) -- (7.7,5.7) node[below] {$b_2$};
    \draw[->,>=stealth,color=red,thick] (6.7,6.) -- (6.1,5.6) node[below] {$b_1$};
    \draw[->,>=stealth,color=red,thick] (6.7,6.) -- (6.6,6.8) node[left] {$e_3$};
    \node[right,color=red,yshift=-0.3cm] (B) at (8.7,6) {$\bU$};
    
\end{tikzpicture}
	\caption{Illustration of a strip $\bss_0$ (with the local basis $\{\bb_s,\bb_q, \be_3\}$) along a curve $\bc$ (with the local basis $\{\btt_0,\bni_0, \bmm_0\}$ in yellow) belonging to  a surface $\bU$ (with the local basis $\{\bb_1,\bb_2, \be_3\}$ in red).}
	\label{Ustrip}
\end{figure}

On the surface $\bU$,  we consider a support curve $\bc \subset \bU $ given by the parametric description $s \to \br_0(s)=\br_\U(u^0(s)) \in \bU$ (here $s\in (0,l)$ is the  arc-length) and we denote by  $\bni_0$ and $\bmm_0$  the  normal and binormal unit vectors (see Figure \ref{Ustrip}). 

Let define now the second order strip  $\bss_0 \subset \bU$  along the curve $\bc$ 
	\begin{equation} \label{sss}
		\bss_0=\{ \br_\U(u(s,q)) \; ; \; u(s,q)=u^0(s)+qv^0(s)+\frac{q^2}{2}w^0(s), s\in (0,l), q\in (-d(s),d(s))\},
	\end{equation} 
	where $2d(s)$ denotes the strip  and $s \to v^0(s)$ and $s \to w^0(s)$ are given by
\begin{equation} \label{w0} 
	v_i^0(s)= (\be_3(u^0(s))\wedge  \btt_0(s))\cdot \bb^i(u^0(s)), \quad	w^0_i(s)=-\frac{1}{2}g^{im}(u^0(s)) \frac{\partial g_{ml}}{\partial u_k}(u^0(s))v^0_k(s)v^0_l(s). 
\end{equation}  

The couple $(s,q)$ defines the curvilinear coordinates of the strip-surface $\bss_0$  with the local basis $\{ \bb_s, \bb_q\}$,  given by  $ \bb_s=\partial_s \br_\U(u(s,q)), \bb_q=\partial_q \br_\U(u(s,q)),$ and the metric tensor $g_{ss}= \vert \bb_s\vert^2,  g_{qq}= \vert \bb_q\vert^2$, $g_{sq}= \bb_s\cdot  \bb_q$.  

Let $\eta\ll 1$ be a small parameter, and let 
\begin{equation}
	k_0^{geo}(s) =k_0(s)\be_3(u^0(s))\cdot \bmm_0(s)
\end{equation} 
denote {\em the geodesic curvature} of the curve $\bc$ on surface $\bss_0$ (see for instance \cite{Spivak99}, chapter 4). One can prove (see Appendix) that if 
\begin{equation}\label{SSK}
	\frac{d(s)}{l}= \Oo(\eta), \quad d(s) \vert \K(u^0(s)) \vert =\Oo(\eta), \quad 	d(s) k_0^{geo}(s)=  \Oo(\eta),
\end{equation} 
then $\bss_0$ is a {\em strip-surface}, i.e.,  the following estimations hold 
\begin{equation}
	\label{SS}
g_{ss}(s,q)-1 + 2qk_0^{geo}(s) = \Oo(\eta^2),\quad g_{qq}
(s,q) - 1 = \Oo(\eta^2),\quad  g_{sq} (s,q)=  \Oo(\eta^2). %\\ \nonumber
\end{equation}

\bigskip

In the local basis $\{\bb^s,\bb^q\}$ of the strip-surface $\bss_0$,  the components $\K_{ss}=\K\bb_s\cdot\bb_s, \K_{qq}=\K\bb_q\cdot\bb_q, \K_{sq}=\K\bb_s\cdot\bb_q$ of curvature tensor $\K$ can be estimated (at first order with respect to $\eta$) as  $\K_{ss}=\K_{ss}^0+q\K_{ss}^1+ \vert\K\vert \Oo(\eta^2)$, $\K_{qq}=\K_{qq}^0+q\K_{q}^1+ \vert\K\vert \Oo(\eta^2)$, $\K_{sq}=\K_{sq}^0+q\K_{sq}^1+ \vert\K\vert \Oo(\eta^2)$ where
\begin{equation} \label{Ksur}
	\K_{ss}^0=\K_{ij}\dot{u}^0_i\dot{u}^0_j, \quad  \K_{qq}^0=\K_{ij}v^0_iv^0_j, \quad    \K_{sq}^0=\K_{ij}\dot{u}^0_iv^0_j, 
\end{equation}
\begin{equation} \label{Ksursq}
	\K_{ss}^1=2\K_{ij}\dot{v}^0_j\dot{u}^0_j, \quad  \K_{qq}^1=2\K_{ij}w^0_jv^0_j, \quad \K_{sq}^1=\K_{ij}(\dot{v}^0_iv^0_j +\dot{u}^0_iw^0_j).
\end{equation}  

\subsubsection{Lagrangian planar strip}
\label{LS}

Let us consider a planar curve $\bCC\subset \RR^2$ given by its parametric description $S \to \bR_0(S) \in \RR^2$ and let $\bTT_0= \frac{d}{d S}\bR_0$ be the tangent vector of the curve $\bCC$. We suppose that $S\in (0,L)$ is the arc-length, hence  $ \vert \bTT_0 \vert =1$ and $L$ is the length of the curve $\bCC$.  We denote by $\bN_0$ and $K_0$ the unit normal and the curvature of $\bCC$ and let 
\begin{equation} 
	\label{SSs}  \bSS_0=\{ \bR(S,Q)=\bR_0(S)+Q\bN_0(S); S\in (0,L), Q\in (-D(S),D(S))\},
\end{equation}
be the planar strip along $\bCC$ where $2D(S)$ is the (local) strip width. The local basis $\{\bb_S,\bb_Q\}$ is given by 
$\bb_S=\frac{\partial }{\partial S} \bR=\bTT_0+Q\frac{d}{d S}\bN_0$, 
$\bb_Q=\frac{\partial }{\partial Q}\bR=\bN_0,$ and from the Frenet formula the metric tensor becomes 
\begin{equation} 
	\label{SSPs}  L_S^2=g_{SS}=1-2QK_0 +Q^2K_0^2,\quad  g_{SQ}=0, \quad  L_Q^2= g_{QQ}=1,
\end{equation}
where $L_S, L_Q$  are the Lam\'e coefficients. The physical basis will be denoted by $\be_S=\bb_S/L_S,  \be_Q= \bb_Q$. 
Since $K_0=K_0^{geo}$,  we see that $\bSS_0$ is a strip-surface (i.e., (\ref{SS}) holds) if and only if 
\begin{equation}\label{SSP}
	\frac{D(S)}{L}=\Oo(\eta),\quad D(S)K_0(S)=\Oo(\eta), \quad \mbox{for all} \; S\in(0,L).
\end{equation}

\bigskip

\begin{figure}[ht]
\centering
\begin{tikzpicture}
    \node[anchor=south west,inner sep=0] at (0,0) 
    {\includegraphics[width=16cm]{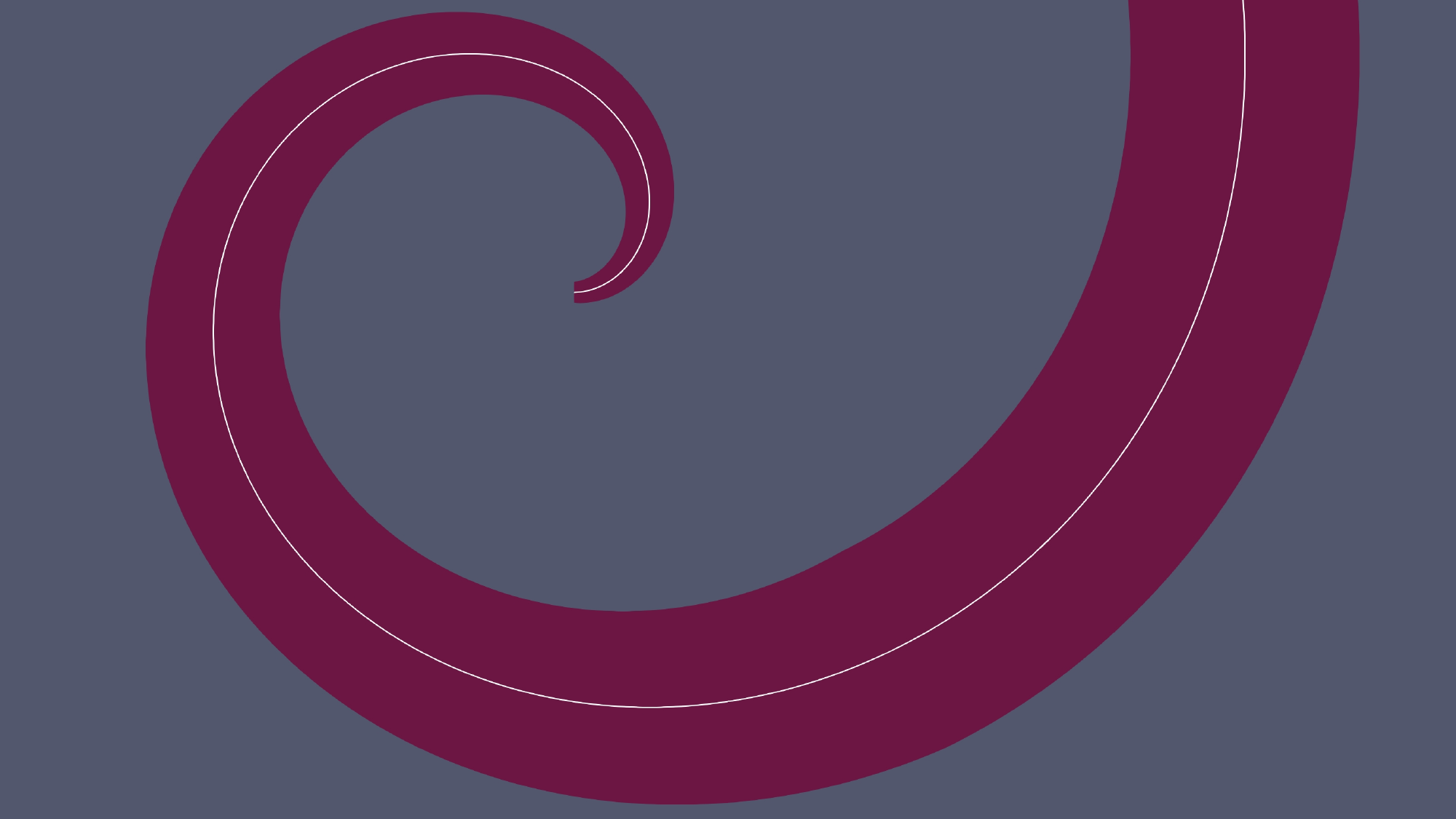}};
    \draw[->,>=stealth,color=white] (8,5) -- (15,5) node[right] {$X_1$};
    \draw[->,>=stealth,color=white] (8,5) -- (8,8.5) node[right] {$X_2$};
    \draw[->,>=stealth,color=yellow,thick] (8,5) -- (10.5,2.25) node[right, yshift=-0.2cm] {$R_0(S)$};
    \draw[->,>=stealth,color=green,thick] (6.9,1.6) -- (6.8,0.6) node[left] {$\be_Q$};
    \draw[->,>=stealth,color=green,thick] (6.9,1.6) -- (5.9,1.7) node[left] {$\be_S$};
    % left part
    \draw[->,>=stealth,color=white,thick] (4.2,3.2) -- (2.,1.) node[right,xshift = 0.2cm] {$Q$};
    \draw[->,>=stealth,color=yellow,thick] (3.6,2.6) -- (2.6,1.6) node[left,yshift=0.3cm] {$N_0$};
    \draw[->,>=stealth,color=yellow,thick] (3.58,2.58) -- (2.6,3.6) node[left] {$T_0$};
    \node[right,color=white,yshift=-0.3cm,xshift=-0.2cm] (A) at (4.2,3.2) {$-D$};
    \node[right,color=white,yshift=-0.3cm] (A) at (3.0,2.0) {$ D$};
    \node[right,color=white,yshift=-0.3cm] (C) at (11.5,4) {$\bCC$};
    \node[right,color=green,yshift=-0.3cm] (C) at (12.7,8) {$\bSS_0$};
\end{tikzpicture}
\caption{Schematic representation of the Lagrangian planar strip $\bSS_0$ (with the local basis $\{\be_S, \be_Q\}$) along a curve $\bCC$ (in white) with its local basis $\{\bTT_0,\bN_0\}$.
}\label{GeomDef}
\end{figure}
\bigskip

\subsubsection{Mapping a planar-strip into a surface-strip} 

Let  $\bx_0 : \bSS_0 \to\bss_0$ be the transformation of the Lagrangian planar strip $\bSS_0$  into the Eulerian strip-surface $\bss_0$  given through the parametric transformation 
\begin{equation} \label{trans}
	s=S, \quad  q=Q, \quad l=L, \quad d=D.
\end{equation}
Then, the  gradient $\bF_0$ acting from the plane of the Lagrangian strip to the 
tangent plane $\bT$ of the surface $\bss_0,$ is given by  
$ \bF_0= \bb_s \otimes\bb^S+\bb_q \otimes\bb^Q,$
and from  (\ref{SS}),(\ref{SSPs}) and (\ref{SSP}) we get $\bF^T_0\bF_0-\bI_2=(1-g_{ss}/g_{SS})\be_S \otimes\be_S$. A straightforward computation gives
$$\bF^T_0\bF_0-\bI_2 =q\left(K_0(s)-k_0^{geo}(s)\right)\be_S \otimes\be_S +\Oo(\eta^2),$$  
\bigskip
and we notice that if we impose the following {\em small-strain curvature  condition} 
\begin{equation} \label{GDS} 
	K_0(s)=k_0^{geo}(s)  + \Oo(\eta^2)
\end{equation}
which relates  the curvature of the Lagrangian curve $\bCC$ to the geodesic curvature of the Eulerian curve $\bc$, then the transformation (\ref{trans})  satisfies  the small strain (or small membrane-strain) assumption for design (\ref{AG}) by choosing $\delta=\eta^2$.  

Let us notice that the above curvature relation (\ref{GDS}) arrives in a different framework in differential geometry: the class of "asymptotic  curves" on a given surface has to have their curvature equal to  the geodesic curvature (see \cite{Spivak99}). The context here is quite different, as we deal with two curves, an  arbitrary  Eulerian one and a plane Lagrangian one.

\subsubsection{Designing the planar strip} 
\label{DesK}

Let us discuss here how to design a Lagrangian planar-strip $\bSS_0$ that can be transformed into a given strip-surface $\bss_0$ under small-strains deformation. 

Here we note that when the curvature $K_0$ is given by (\ref{GDS}), it is straightforward to obtain the planar curve $\bCC.$ If $\Theta(s)$ is such that
$\bTT_0(s)=\cos(\Theta(s))\bc_1+\sin(\Theta(s))\bc_2$,  
we obtain 
\begin{equation} 
	\label{The}\Theta(s)=\Theta_0+\int_0^s K_0(t)\; dt, \quad \bR_0(s)=\int_0^s  \bTT_0(s)\; dt,
\end{equation} 
which is the parametric description of the support curve $\bCC$.

By choosing the width $d=D$ such that 
\begin{equation} \label{D} 
	d(s)K_0(s)=\Oo(\eta),
\end{equation} 
the designed planar-strip $\bSS_0$, given by (\ref{SSs}), admits a small-strain transformation of order $\Oo(\eta^2)$ into the strip-surface $\bss_0$, i.e., (\ref{AG}) holds.

\subsection{Designing the plate-strip for a given shell-strip} 

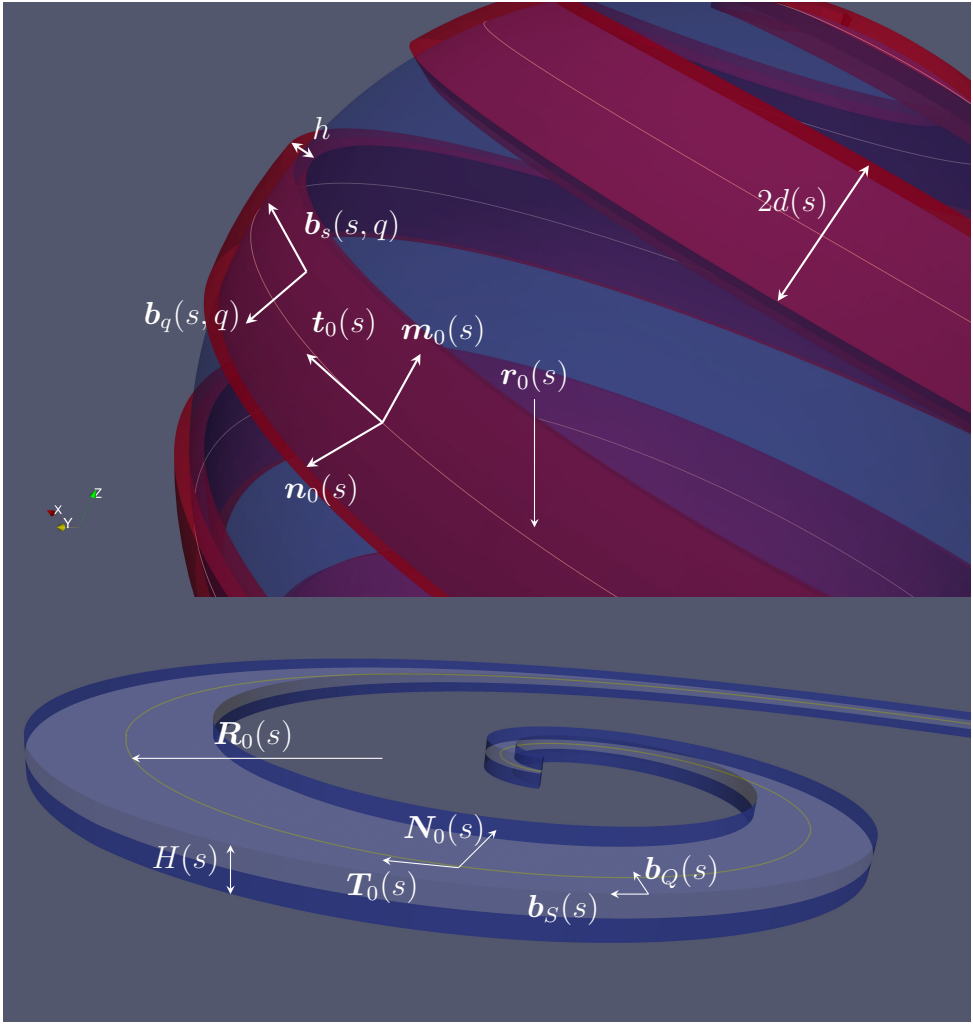
\begin{figure}[th!]
\centering
\begin{tikzpicture}
    \node[anchor=south west,inner sep=0] at (0,0) {\includegraphics[width=0.8\textwidth]{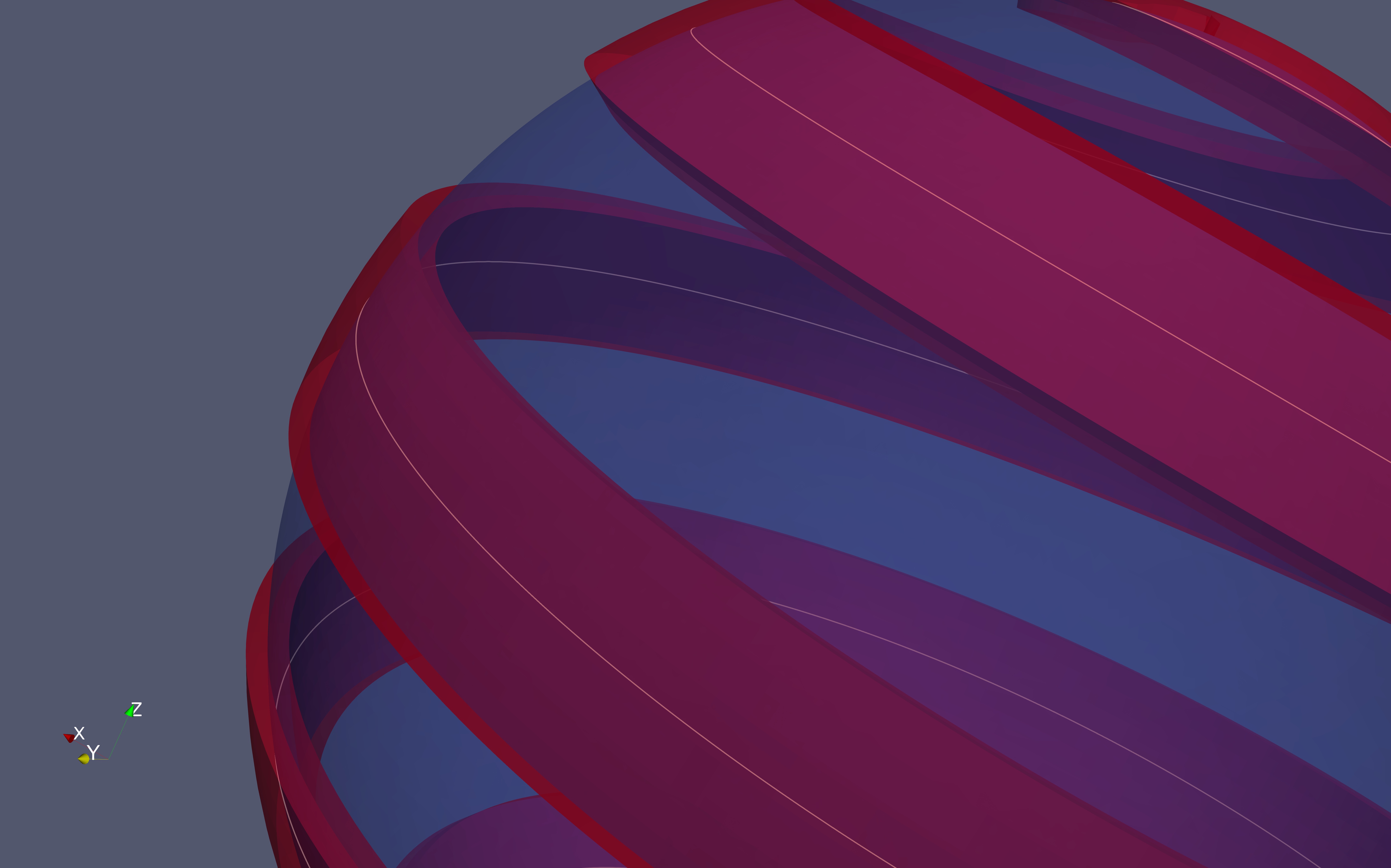}};
    %\draw[red,ultra thick,rounded corners] (7.5,5.3) rectangle (9.4,6.2);
    % ===  r_0 =====
    \node[text=white] (A) at (7,3) {$\br_0(s)$};
    \draw[->,>=stealth,color=white,thin] (7,2.7) -- (7,1);
    % === frame on the curve ===
    \draw[->,>=stealth,color=white,thick] (5,2.39) -- (4,3.3);
    \node[text=white] (A) at (4.5,3.7) {$\btt_0(s)$};
    \draw[->,>=stealth,color=white,thick] (5,2.39) -- (4,1.8);
    \node[text=white] (A) at (4.2,1.5) {$\bni_0(s)$};
    \draw[->,>=stealth,color=white,thick] (5,2.39) -- (5.5,3.3);
    \node[text=white] (A) at (5.8,3.6) {$\bmm_0(s)$};
    %=== local basis on the surface
    \draw[->,>=stealth,color=white,thick] (4,4.39) -- (3.5,5.3);
    \node[text=white] (A) at (4.6,5) {$\bb_s(s,q)$};
    \draw[->,>=stealth,color=white,thick] (4,4.39) -- (3.2,3.7);
    \node[text=white] (A) at (2.5,3.8) {$\bb_q(s,q)$};
    %=== thickness ===
    \draw[<->,>=stealth,color=white,thick] (4.1,5.9) -- (3.8,6.1);
    \node[text=white] (A) at (4.2,6.3) {$h$};
    %==== largeur ===
    \draw[<->,>=stealth,color=white,thick] (10.2,4) -- (11.4,5.8);
    \node[text=white] (A) at (10.4,5.3) {$2d(s)$};
\end{tikzpicture}

\vspace*{-0.1cm}
\begin{tikzpicture}
    \node[anchor=south west,inner sep=0] at (0,0) {\includegraphics[width=0.8\textwidth]{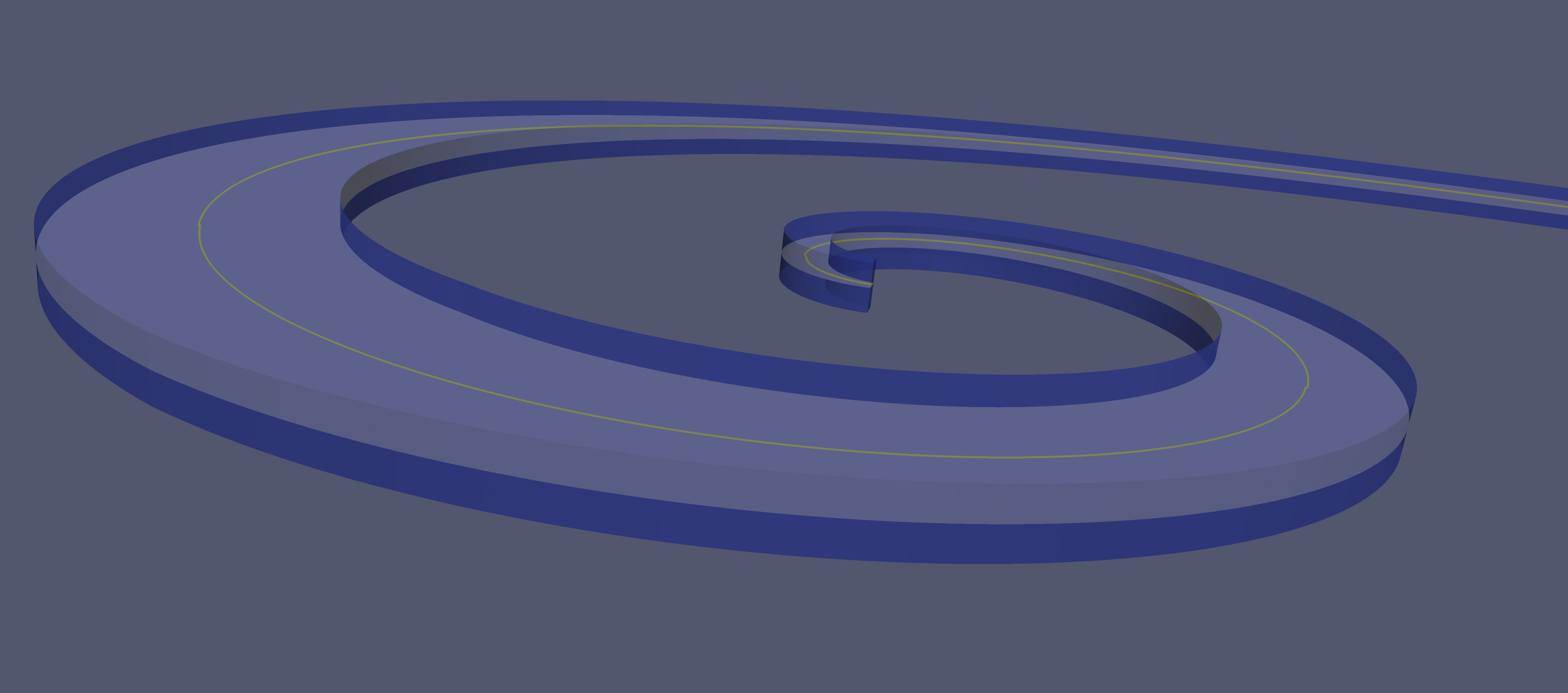}};
    \draw[<-,>=stealth,color=white,thin] (1.7,3.5) -- (5,3.5);
    \node[text=white] (A) at (3.3,3.8) {$\bR_0(s)$};
    \draw[->,>=stealth,color=white,thin] (6,2.05) -- (5,2.15) node[below] {$\bTT_0(s)$};
    \draw[->,>=stealth,color=white,thin] (6,2.05) -- (6.5,2.55) node[left] {$\bN_0(s)$};
    \draw[<->,>=stealth,color=white,thin] (3,1.7) -- (3,2.35) node[left,yshift=-0.2cm] {$H(s)$};
    \draw[->,>=stealth,color=white,thin] (8.5,1.7) -- (8.,1.7) node[left,yshift=-0.2cm] {$\bb_S(s)$};
    \draw[->,>=stealth,color=white,thin] (8.5,1.7) -- (8.3,2.) node[right,yshift=-0.0cm] {$\bb_Q(s)$};

\end{tikzpicture}
\caption{The geometric elements of an Eulerian shell-strip $\bss$ constructed  from  a mid-surface-strip $\bss_0$ along a support curve $\bc$ (top picture) and the geometric elements of a Lagrangian plate-strip $\bSS$ constructed from  a planar  mid-surface-strip $\bSS_0$ along a support curve $\bCC$ (bottom picture).}
\label{fig:StripShell}
\end{figure}

Let $\bss$ be an Eulerian shell-strip constructed (see (\ref{ssE}))
from a surface-strip $\bss_0$ (see Figure \ref{fig:StripShell}) and let the Lagrangian plate-strip $\bSS$  be constructed (see (\ref{SSL})) from a planar-strip $\bSS_0$ with  $h=H(1+\Oo(\delta^2))$. If the planar-strip $\bSS_0$ is designed such that (\ref{GDS}) and (\ref{D}) hold, then the transformation $\bx_0:\bSS_0\to  \bss_0$ given by (\ref{trans}) has a small-strain  deformation (\ref{AG}) of order $\eta^2$. If we now choose the thickness $h$ to be of order $\eta=\sqrt{\delta}$  with respect to the strip width $d$, i.e.,  
\begin{equation} \label{Horder}
	\delta=\eta^2, \quad 	\frac{h}{d}=\Oo(\eta),
\end{equation}
then from  (\ref{SSK})
we obtain (\ref{H}). This means that if (\ref{GDS}) and (\ref{D}-\ref{Horder}) are satisfied, this corresponds precisely to the case of plate-to-shell design theory developed in \cite{danescu2020shell} and briefly presented in the section 2.

\bigskip 

The designed pre-stress moment $\bM^*$ can be obtained from (\ref{M0}) and from the expression of the Lagrangian curvature $\bK$ given by 
\begin{equation}\label{K0}
	\bK=\frac{\K\bb_s\cdot \bb_s}{L_S^2}\be_S\otimes\be_S+ \frac{\K\bb_s\cdot  \bb_q}{L_S}(\be_Q\otimes\be_S+\be_S\otimes\be_Q)+\K\bb_q\cdot \bb_q\be_Q\otimes\be_Q ).
\end{equation}

\bigskip 

One  can develop  the  components of the Lagrangian  curvature tensor $\bK$ in the physical basis $\{\be_S,\be_Q\}$ as  $K_{SS}=K_{SS}^0+qK_{SS}^1+ \vert\K\vert \Oo(\eta^2)$, $K_{QQ}=K_{QQ}^0+qK_{QQ}^1+ \vert\K\vert \Oo(\eta^2)$, $K_{SQ}=\K_{SQ}^0+q\K_{SQ}^1+ \vert\K\vert \Oo(\eta^2)$.   Having in mind  (\ref{Ksur})-(\ref{Ksursq}), we get:
\begin{equation}\label{K0sur}
	K_{SS}^0=\K_{ss}^0,\qquad K_{SQ}^0=\K_{sq}^0,\qquad K_{QQ}^0=\K_{qq}^0,\end{equation}
\begin{equation}
	K_{SS}^1=\K_{ss}^1+2K_0,\qquad K_{SQ}^1=\K_{sq}^1+K_0,\qquad K_{QQ}^1=\K_{qq}^1. 
\end{equation}

\section{Applictions to non-developable shells}  

In this section, we illustrate the plate-to-shell design equations introduced in the previous section, in order to design a sphere and a torus by using only isotropic materials with a weak heterogeneity, i.e., for which (\ref{Isotrop}) and (\ref{IsoWH}) hold.

\subsection{Strips on spherical surfaces}

Let $(r, \theta, \phi)$ be the spherical coordinates in the Eulerian description and denote by  $\be_r=\be_r(\theta,\phi), \be_\theta=\be_\theta(\theta,\phi), \be_\phi= \be_\phi(\phi)$ the local physical basis in the Eulerian description.
We consider the  spherical surface  $ \bU=\{r=R_*\} $ of radius $R_*$ with Lam\'e coefficients  $L_\theta=R_*, L_\phi=R_*\sin(\theta)$ and the unit normal   $\be_3(\theta, \phi)=\be_r(\theta,\phi)$,  while the curvature tensor is $\K=-\frac{1}{R_*}\left(\be_\theta\otimes\be_\theta+ \be_\phi\otimes\be_\phi\right).$

On  $ \bU$  we consider a  curve  $\bc \subset \bU $  given by its parametric  description $s \to \br_0(s)=R_*\be_r(\theta^0(s),\phi^0(s))  \in \bU$.  
The tangent vector is  $\btt_0(s)=R_*\left(\dot{\theta}^0\be_\theta+\sin(\theta^0)\dot{\phi}^0\be_\phi\right),$  
and $s\in (0,l)$ is the arc-length, hence  $\theta^0, \phi^0$ and $s$ are related by 
$$ R_*^2\left((\dot{\theta}^0(s))^2 +\sin^2(\theta^0(s))(\dot{\phi}^0(s))^2\right)=1.$$  
A straightforward computation gives $ \bni_0^\U(s)=\be_r(\theta^0,\phi^0)  \wedge  \btt_0=R_*\left(-\sin(\theta^0)\dot{\phi}^0\be_\theta+\dot{\theta}^0\be_\phi \right)$ so that
$v^0_\phi=\dot{\theta}^0/\sin(\theta^0),   
v^0_\theta=-\sin(\theta^0)\dot{\phi}^0$ and 
$w^0_\phi=\cot(\theta^0)\dot{\theta}^0\dot{\phi}^0,   
w^0_\theta=0$ and finally, from (\ref{w0}), 
we can construct (see subsection \ref{Surface}) the strip-surface $\bss_0$:
$$\bss_0=\{\phi=\phi^0+q\frac{\dot{\theta}^0}{\sin(\theta^0)}+\frac{q^2}{2}\cot(\theta^0)\dot{\theta}^0\dot{\phi}^0, \; \theta=\theta^0-q\sin(\theta^0)\dot{\phi}^0 \; ; \; s\in(0,l), q\in(-d(s),d(s))\}.$$
Then (\ref{SSK}) reads 
\begin{equation}\label{dRsphere}
	\frac{d(s)}{R_*}=\Oo(\eta), \quad 	d(s)k_0^{geo}(s)=\Oo(\eta),   
\end{equation}
where the geodesic curvature is given by 
\begin{equation}\label{dRR}
	k_0^{geo}=R_*^2\left( \sin(\theta^0)\dot{\theta}^0\ddot{\phi}^0-[\ddot{\theta}^0-
	\sin(\theta^0)\cos(\theta^0)(\dot{\phi}^0)^2]\sin(\theta^0)\dot{\phi}^0 +
	2(\dot{\theta}^0)^2\dot{\phi}^0\cos(\theta^0) \right).
\end{equation}

Let $\bCC$ be the designed planar curve with geodesic curvature $K_0(s)=k_0^{geo}(s)$ given by the above formula, and let $\bSS_0$ be the planar-strip designed on the support Lagrangian curve $\bCC$ with the width $d(s)$ such that (\ref{D}) holds. Then, from the small-strain membrane condition (\ref{AG}) we get $\displaystyle \bK=\frac{1}{R_*}( \bI_2+\Oo(\delta))$.  

\bigskip

To resume, we find that {\em a strip  $\bss$ of a spherical shell of radius $R_*$ along the curve $\bc$, could  be designed from a plate-strip $\bSS$  along a curve $\bCC$  if  
	(\ref{dRsphere}) holds and the Lagrangian curvature $K_0=k_0^{geo}$ is given by  (\ref{dRR}). The pre-stress moment in $\bSS_0$ is given by
	$\displaystyle  \bM^*=-\frac{H^3}{12R_*}\bMM_2\bI_2$  
	and can be obtained with an isotropic and homogeneous pre-stress, i.e., $\bS^*_2=\sigma^* \bI_2$, where }
\begin{equation}\label{RRs}
	\hat{\sigma}^*=\frac{H}{12R_*}
	(\bar{C}_{11}^2+\bar{C}_{12}\bar{C}_{11}-2\bar{C}_{12}^2)/\bar{C}_{11}=
	\frac{H}{12R_*} \frac{2\bar{\mu}(3\bar{\lambda}+2\bar{\mu})}{\bar{\lambda}+2\bar{\mu}}.
\end{equation}
Here, for simplicity, we used $\bar{C}_{11} = \bar{\lambda}+2\bar{\mu}$ and $\bar{C}_{12}=\bar{\lambda}.$

\subsubsection{"Orange-peeling"  strips}

In the particular case when $\ddot{\phi}^0 =0$ (i.e., $\phi^0(s)=As+B$), the normalization condition reads 
$(\dot\theta^0)^2(s) + A^2\sin^2(\theta^0(s)) =1/R_\star^2$, while the  expression of the Lagrangian curvature now has a simpler form. Relations (\ref{dRsphere})-(\ref{dRR}) become
\begin{equation}\label{KspereS}
	k_0^{geo}(s)=K_0(s)= A\cos\theta^0(s),\quad 	d(s)A\cos\theta^0(s)=\Oo(\eta), \quad 	\frac{d(s)}{R_*}=\Oo(\eta).
\end{equation}
In Figure \ref{Ustrip}, we have plotted a spherical strip  with  the width computed  from the above formula  
$
\displaystyle d(s)=\eta \min\{R_*, \frac{1}{A\cos\theta^0(s)}\},
$
with $d(s)K_0(s)=d(s)A\cos\theta^0(s)$ in color scale.  We note that the strip width has to be drastically reduced when the spherical strip is near the poles. 

\begin{figure}[ht!]
	\centering
\includegraphics[width=16cm]{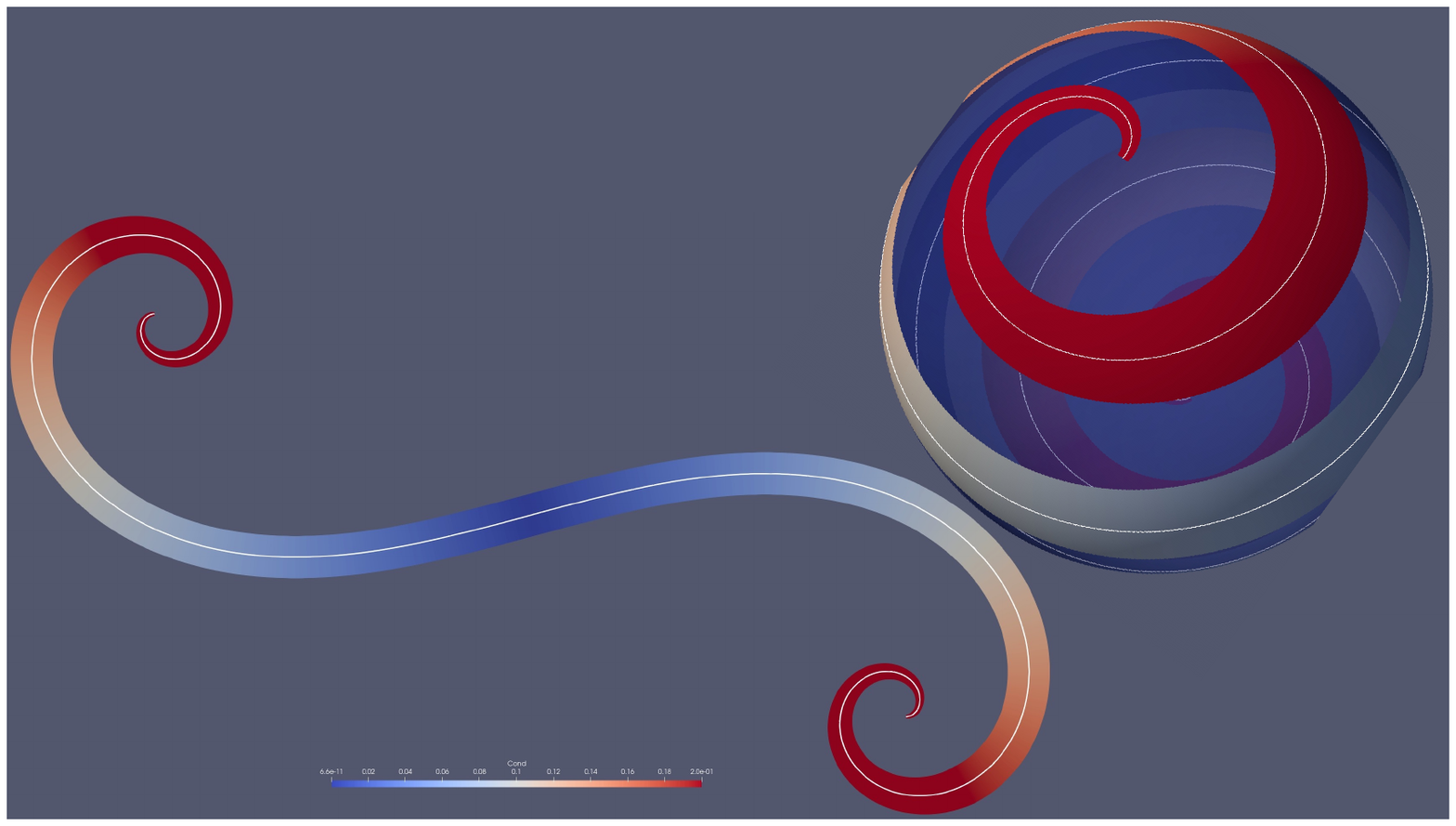}

\vspace*{-1cm}
	\caption{Right: an "orange peeling"  spherical strip with $\ddot{\phi}^0 =0$. Left:   the designed planar-strip computed for $\eta=0.2, \delta=4\%$ with $K_0d=k_0^{geo}d$ in color scale. The Lagrangian and the Eulerian configurations are plotted at different length scales.}
	\label{Sphere_strip}
\end{figure}

\subsubsection{Covering a sphere with meridian strips}

\begin{figure}[ht!]
\centering
\includegraphics[width=7.5cm]{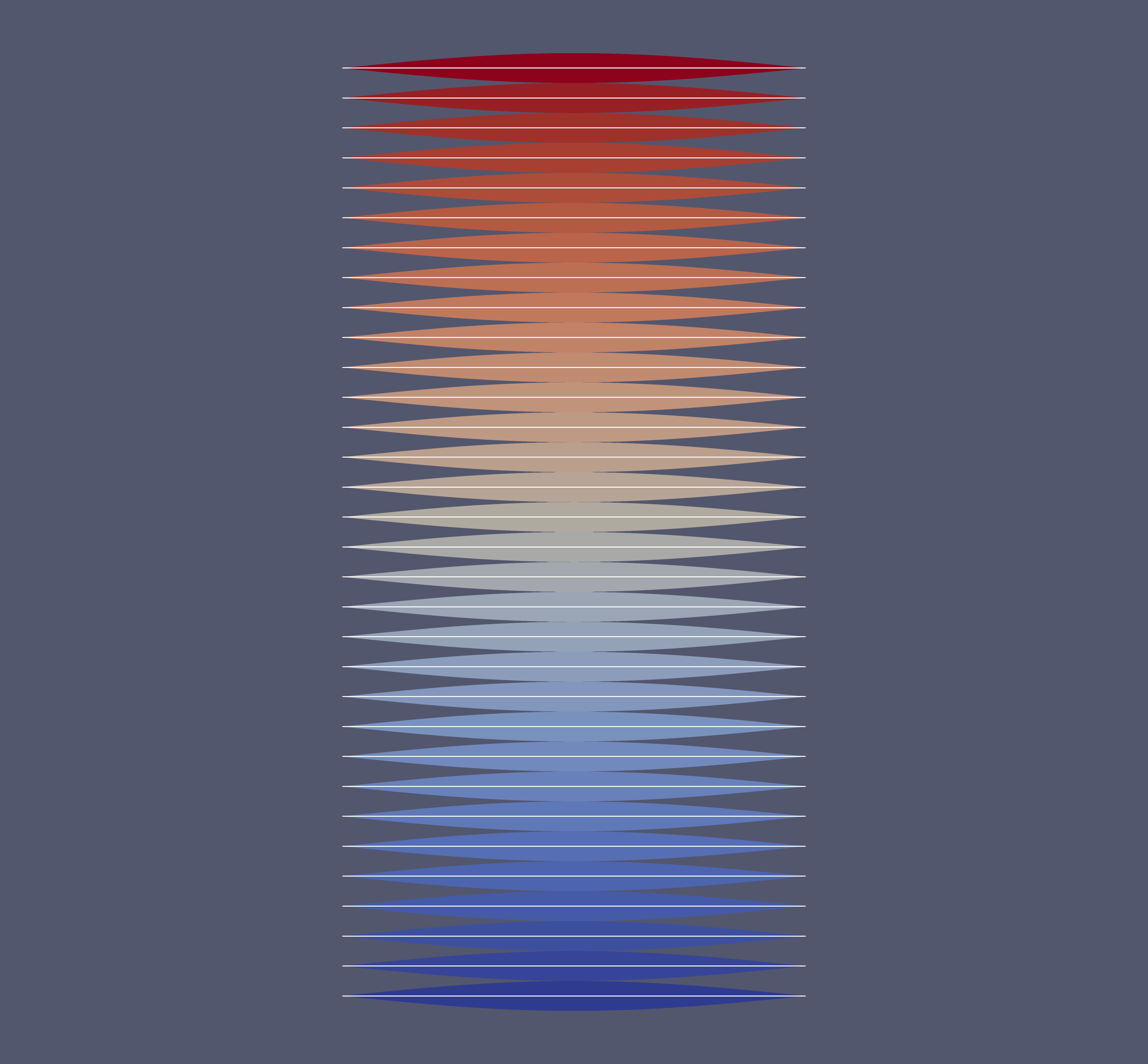}  \hspace{-0.5cm}
\includegraphics[width=7.5cm]{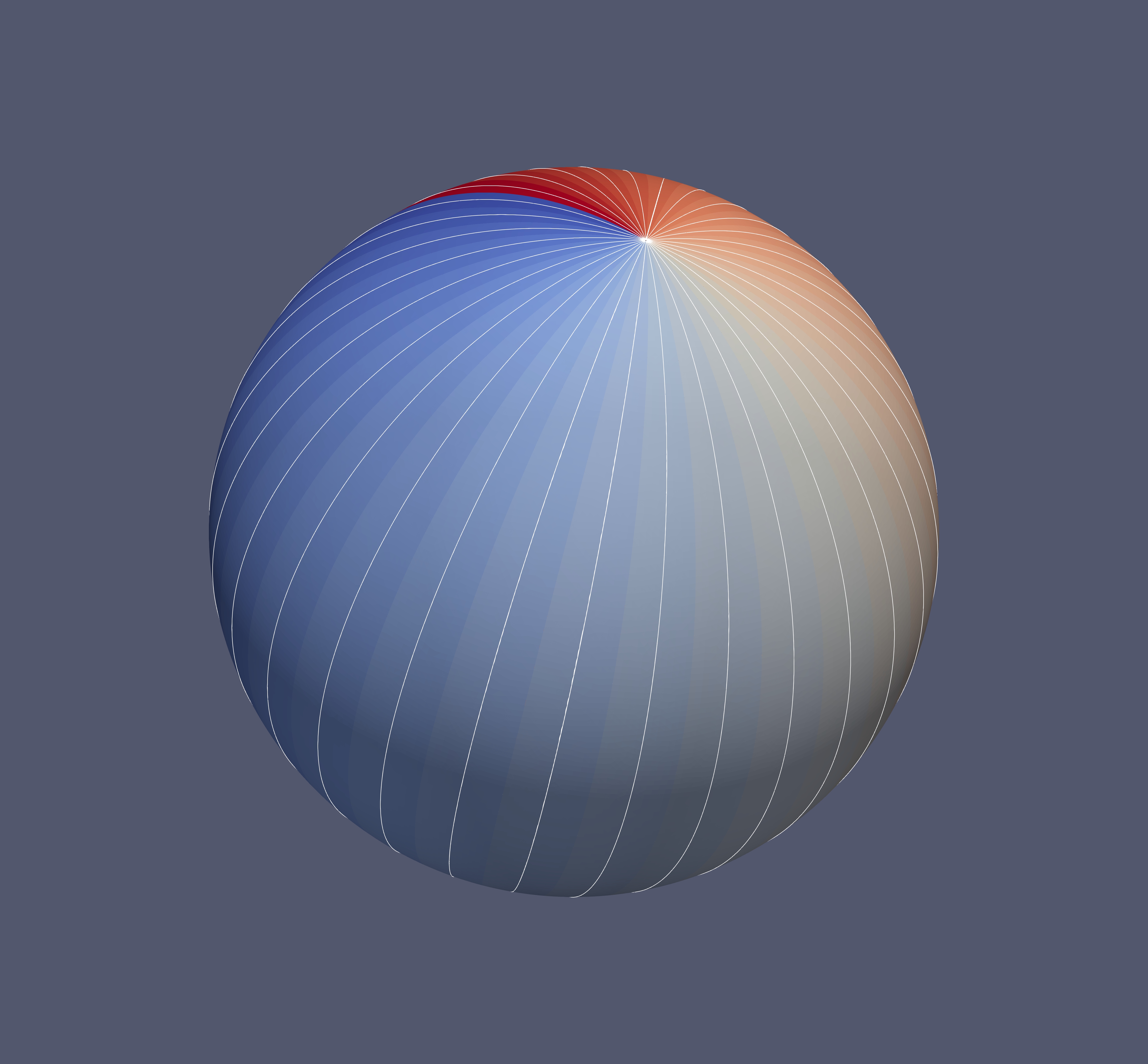}
\caption{Right: a sphere covered by 32  meridians strips (with strip index in color scale), corresponding to a deformation of $\delta=1\%$. Left:  the designed Lagragian configuration. The Lagrangian and the Eulerian configurations are plotted at different length scales.}
\label{SphereMeridians}
\end{figure}

For meridian curves, i.e.,  $\theta^0=s/R_*, \phi^0=const$,  we get $K_0=0$, hence  the designed planar strip is a straight strip with  $d(s)\leq \eta R_*$.  In order to  look for conditions that ensure the complete covering of  the sphere, 
we define the meridian strips $$\bss_0^k=\{\phi=\phi_k+\frac{q}{\sin(s/R_*)}, \; \theta=S/R_*+\pi/2\; ; \; s\in(-\pi R_*/2,\pi R_*/2), q\in(-d(s),d(s))\}.$$
Let $N^{mer}=[\pi/\eta] +1$ (we have denoted by $[x]=\max\{n \; ; \;  n\leq x\} $ the entire part of $x$) be the number of meridians and from the covering condition $\phi_k+d(s)/\sin(S/R_*)=\phi_{k+1}-d(s)/\sin(s/R_*)$,  we obtain 
$$\phi_k=\frac{2\pi}{N^{mer}}k, \quad k=0,1,...,N^{mer}-1, \quad \quad  d(s)=\frac{\pi R_*}{N^{mer}}\sin(s/R_*).$$ 
We note that for a small deformation of order of  $\delta=1\%$  (i.e., $\eta=0.1$),  we need at least $N^{mer}=32$  meridian strips. More precisely, for 32 meridians we obtain an approximation of $\eta=\pi/32\approx 0.0981, \delta=\pi^2/1024\approx 0.963\%$ (see Figure \ref{SphereMeridians} for a graphical illustration). 
  
\subsubsection{Covering a sphere with parallel strips}

For constant latitude curves, i.e.,  $ \theta^0=const, \phi^0=s/(R_*\sin(\theta_0))$, 
we have, in addition to (\ref{dRsphere}),  $d\cot(\theta^0)/R_*=\Oo(\eta)$ and $K_0(s)=\cot(\theta^0)/R_*$.  To design constant latitude strips,  let us denote by  $$\bss_0^k=\{\phi=\frac{s}{R_*\cos(\theta_k) } \;  ; \; \theta=\frac{\pi}{2}-\theta_k-\frac{q}{R_*}\; ; \; s\in(-\pi R_*\cos(\theta_k),\pi R_*\cos(\theta_k)), q\in(-d_k,d_k)\}$$
and let $N^{lat}$ be the number of strips needed to partially cover the sphere, for $\theta \in (\eta, \pi-\eta)$. The remaining parts, the spherical callus $0< \theta \leq \eta$ and $\pi-\eta\leq \theta<\pi$ can be covered from a planar disc with small strain (see \cite{danescu2020shell}).

\begin{figure}[ht!]
	\centering
	\includegraphics[width=16cm]{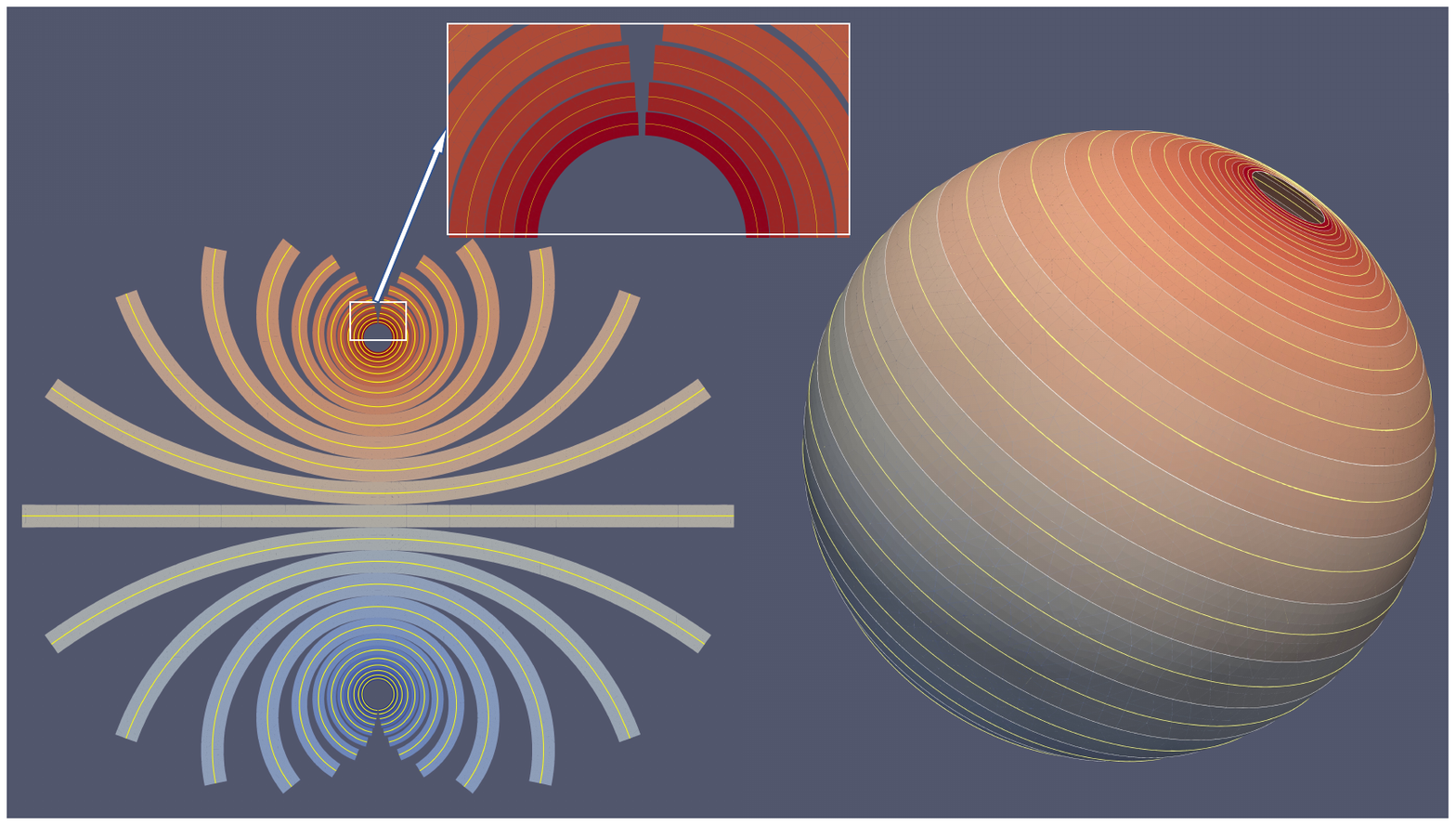}
	\vspace{-1.5cm}
	\caption{Right: an optimized covering  of an almost complete sphere ($\eta<\theta<\pi-\eta$) with 25 parallel strips (with strip index in color scale), corresponding to a deformation of $\eta^2=\delta=1\%$. Left: the designed Lagrangian configuration. The strip positions and widths were computed from (\ref{DSphereOp}-\ref{DSphereOG}). The Lagrangian and the Eulerian configurations are plotted at different length scales.
	}
	\label{SphereParallels}
\end{figure}

For a symmetric solution,  let us consider $N^{lat}=2M+1$, with $k=-M,...,0,...M$ and $\theta_0=0,  \theta_{-k}=-\theta_{k}$. From (\ref{KspereS}) and the covering condition,  we obtain the following recursive system $$d_k\leq\eta R_* \min\{1, \cot(\theta_k)\}, \quad \theta_{k+1}-\theta_{k}= \frac{d_k+d_{k+1}}{ R_* }.$$
For $ \theta\leq\pi/4$ we can consider constant-width strips 
\begin{equation}\label{DSphereOp}
	\theta_k=2k\eta, \quad  d_k=\eta R_*, \quad \mbox{for}\;  \vert k \vert \leq M_0=[\frac{\pi}{8\eta}-\frac{1}{2}]\leq M,
\end{equation}
while for $k>M_0$,  we have to solve recursively the nonlinear equation $f(x)=x-\eta\cot(x)=\theta_{k}+d_k/R_*$ to find $\theta_{k+1} =f^{-1}(\theta_{k}+ d_k/R_*)$, i.e., 
\begin{equation}\label{DSphereOG} 
	f(\theta_{k+1})=\theta_{k}+ \frac{d_k}{ R_* }, \quad d_{k+1}=R_*(\theta_{k+1}-\theta_{k}) -d_k, \quad \mbox{for}\;  M_0 <  \vert k \vert \leq M.
\end{equation}

The strip positions $\theta_k$ and strip widths $d_k$ can be computed from the iterative system (\ref{DSphereOp}-\ref{DSphereOG}),  while the number of strips $N^{lat}$ is computed such that $\theta_M>\pi/2-\eta$.  In Figure \ref{SphereParallels}, we illustrate the implementation of the above optimal design procedure of a sphere for a given strain 
$\delta=1\%$ ($\eta=0.1$) corresponding to $N^{lat}=25$ strips. 

From a technological prespective, the sharp angles between neighbour strips along the vertical line in Figure \ref{SphereParallels} are very difficult to realize since they are at the lower limit of the photolythography. However, based on a simplified version of the design in Figure \ref{SphereParallels} we realized in \cite{arxiv} an exprimental proof of the design of the sphere.

\subsection{Strips on a torus }

To explore beyond objects with constant curvature, we further discuss the shell-strip covering of the torus. Let $(r, \phi, z)$ be the cylindrical coordinates and denote by  $\be_r=\be_r(\phi), \be_\phi=\be_\phi(\phi), \be_z$ the local physical basis. 

We consider the torus $\bU$ with radii $R_*>r_*,$ given by the parametric description  
$(\phi,\psi)\to \br_{\bU}(\phi,\psi)=(R_*+r_*\cos\psi)\be_r(\phi)+ r_*\sin\psi\be_z$.    The local basis is $\bb_\phi=(R_*+r\cos\psi)\be_\phi, \bb_\psi=r_*(-\sin\psi\be_r(\phi)+\cos\psi\be_z)$, while the Lam\'e coefficients are  $L_\psi=r_*, L_\phi=R_*+r_*\cos\psi$. We can compute the physical basis  $\be_\phi=\be_\phi(\phi), \be_\psi=\be_\psi(\phi,\psi)=-\sin\psi\be_r(\phi)+\cos\psi\be_z$, the unit normal   $\be_3(\phi,\psi)=\cos\psi\be_r(\phi)+\sin\psi\be_z$ and the curvature tensor  $$\K=-\frac{1}{r_*}\be_\psi\otimes\be_\psi-\frac{\cos\psi}{R_*+r_*\cos\psi} \be_\phi\otimes\be_\phi.$$

On $ \bU$ we consider a curve $\bc \subset \bU $ given by its parametric 
description $s \to \br_0(s)=\br_{\bU}(\phi^0(s),\psi^0(s))  \in \bU$, where $s\in (0,l)$ is the arc-length.  
The tangent vector is  
$\btt_0(s)=r_*\dot{\psi}^0\be_\psi(\phi^0,\psi^0)+\dot{\phi}^0(R_*+r_*\cos\psi^0)\be_\phi(\phi^0)$  and $\phi^0,\psi^0,$ and $s$ are related by 
\begin{equation}\label{NormTore} 
	r_*^2(\dot{\psi}^0(s))^2 +(\dot{\phi}^0(s))^2(R_*+r_*\cos\psi^0(s))^2 =1.
\end{equation} 
We can now compute
$$ \bni_0^\U(s)=\be_3(\phi^0,\psi^0)  \wedge  \btt_0=-r_*\dot{\psi}^0\be_\phi(\phi^0)+\dot{\phi}^0(R_*+r_*\cos\psi^0(s)) \be_\psi(\phi^0(s),\psi^0(s))$$  
which gives $v^0_\phi=-r_*\dot{\psi}^0/(R_*+r_*\cos\psi^0),   
v^0_\psi=\dot{\phi}^0(R_*+r_*\cos\psi^0)/r_*$ from  (\ref{w0}). After some additional computations we get 
$$w^0_\phi=-r_*\sin\psi^0/(R_*+r_*\cos\psi^0) \dot{\phi}^0\dot{\psi}^0, \;   
w^0_\psi=0$$
so that (see subsection \ref{Surface}) the strip-surface $\bss_0$ is obtained as
$$\bss_0=\{\br_{\bU}(\phi,\psi) \; ; \; \phi=\phi^0+qv^0_\phi+\frac{q^2}{2}w^0_\phi, \;   \psi=\psi^0+qv^0_\psi \; ; \; s\in(0,l), q\in(-d(s),d(s))\}.$$
Let us compute the geodesic curvature: bearing in mind that $\dot{\be}_\phi=\dot{\phi}(\sin\psi\be_\psi-\cos\psi\be_3)$ and $\dot{\be}_\psi=-\dot{\phi}\sin\psi\be_\phi-\dot{\psi}\be_3$,  we get $\dot{\btt}_0(s)=\left(\ddot{\phi}_0(R_*+r_*\cos\psi_0)-2\dot{\phi_0}\dot{\psi}_0r_*\sin\psi_0\right)\be_\phi+ \left(r_*\ddot{\psi}_0-\dot{\phi}^2_0(R_*+r_*\cos\psi_0) \right)\be_\psi-\left(r_*\dot{\psi}^2_0 +\dot{\phi}^2_0(R_*+r_*\cos\psi) \right)\be_3$ and since $k_0^{geo} =\be_3\cdot (\btt_0\wedge \dot{\btt}_0)$,  we obtain the geodesic curvature as
$$k_0^{geo}=(R_*+r_*\cos\psi_0)\left(r_*(\ddot{\psi}_0\dot{\phi}_0-\ddot{\phi}_0\dot{\psi}_0)+\dot{\phi}^3_0(R_*+r_*\cos\psi_0)\sin\psi_0\right)+2\dot{\phi}_0\dot{\psi}^2_0r_*^2\sin\psi_0.$$
Then (\ref{SSK})-(\ref{SS}) reads 
\begin{equation}\label{dTore} 
	d(s)\max\{\frac{1}{r_*}, \frac{\cos\psi_0(s)}{R_*+r_*\cos\psi_0(s)}\}=\Oo(\eta), \quad 	d(s)k_0^{geo}(s)=\Oo(\eta).   
\end{equation}

Let $\bCC$ be the designed planar curve with the curvature $K_0(s)=k_0^{geo}(s)$, given by the above formula, and let  the planar-strip $\bSS_0$ be designed on the support Lagrangian curve $\bCC$ with the width $d$ such that  (\ref{dTore}) holds. Then the small-strain membrane condition (\ref{AG}) holds.  

The Lagrangian curvature tensor   $\bK$ can be computed from (\ref{K0}) and the formulae of the curvature tensor 
$\K\bb_s\cdot \bb_s=\cos\psi(R_*+r_*\cos\psi)(\partial_s\phi)^2+r_*(\partial_s\psi)^2, \; \K\bb_s\cdot \bb_q=\cos\psi(R_*+r_*\cos\psi)\partial_s\phi\partial_q\phi+r_*\partial_s\psi\partial_q\psi, \; \K\bb_q\cdot \bb_q=\cos\psi(R_*+r_*\cos\psi)(\partial_q\phi)^2+r_*(\partial_q\psi)^2$. Now using  (\ref{K0sur}), we get at order $\eta^0$:  
$$K_{SS}^0=-\frac{1}{r_*} +\frac{\dot{\phi}_0^2(R_*+r_*\cos\psi)R_*}{r_*}, \; K_{QQ}^0=-\frac{1}{r_*} +\frac{\dot{\psi}_0^2R_*r_*}{R_*+r_*\cos\psi},\;  K_{SQ}^0=\dot{\phi}_0\dot{\psi}_0R_*.$$  

\bigskip

\begin{figure}[ht!] 
	\centering
	\includegraphics[width=16cm]{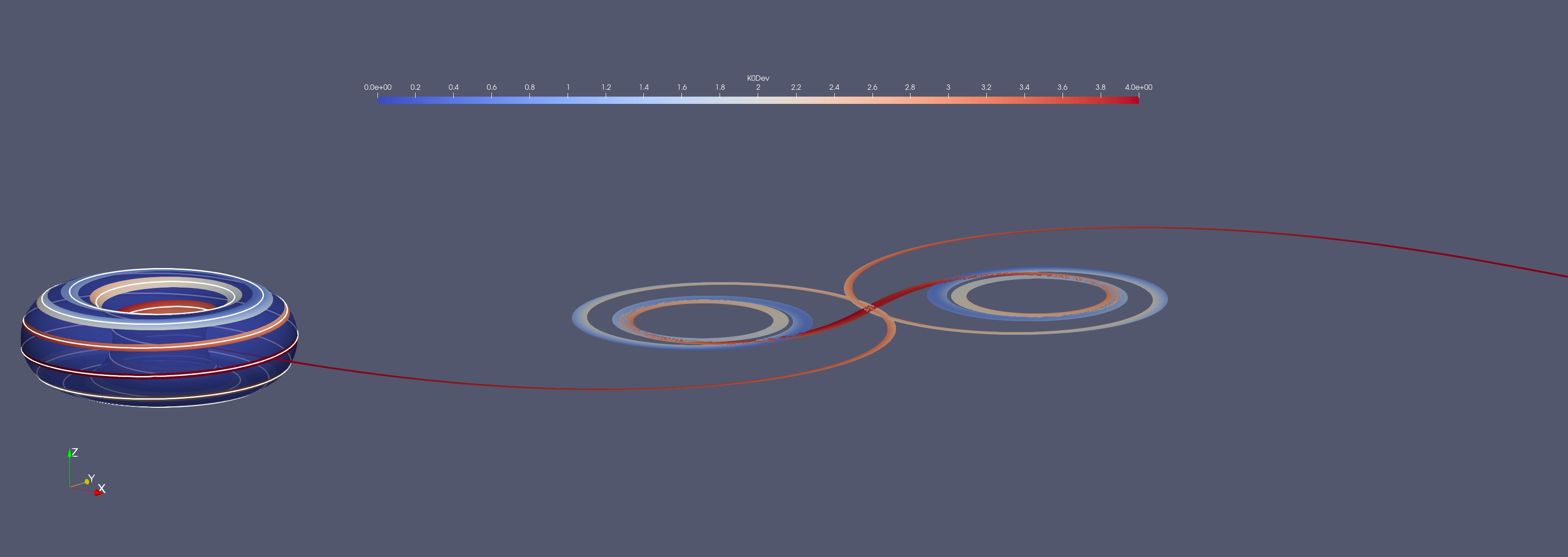}
	\caption{The Eulerian (left) and Lagrangian (right) descriptions of a  closed parallel strip on a torus with the width $d$ given by  (\ref{dTore}) for $\eta^2=\delta=1\%$ (with $\vert\bK^D\vert$ in color scale).}
	\label{Thorus1}
\end{figure}

\begin{figure}[th!] 
		\centering
	\includegraphics[width=16cm]{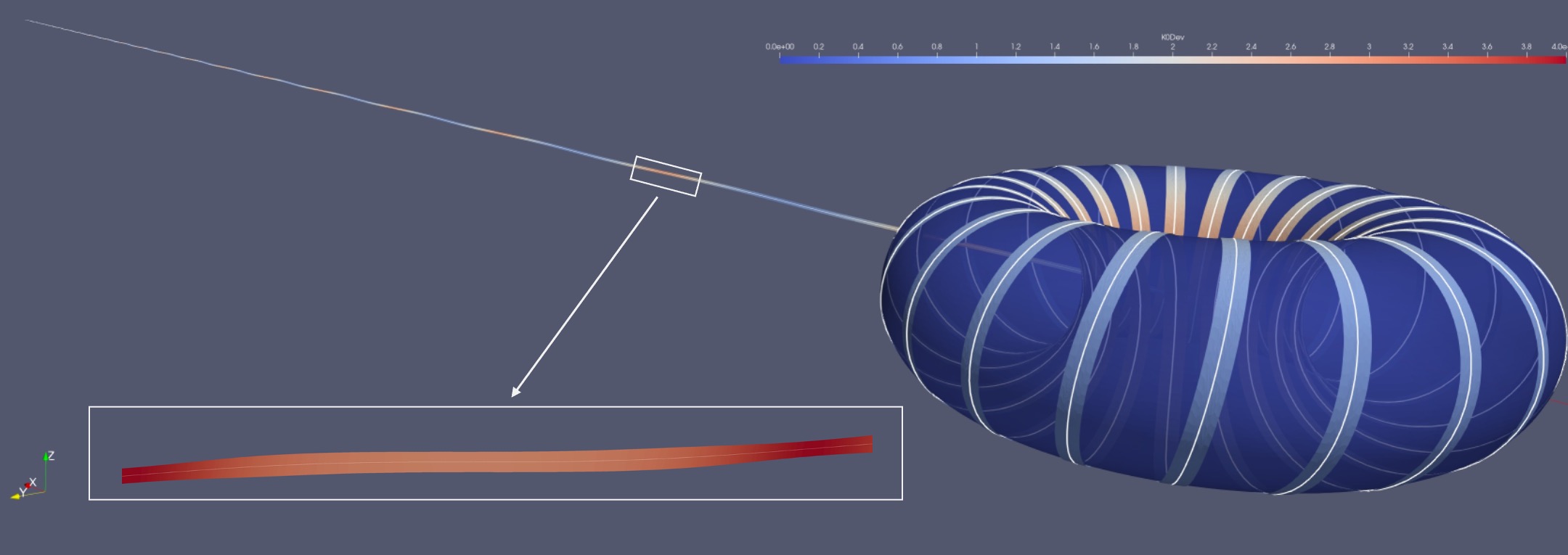}	
		%\vspace{-2cm}
		\caption{The Eulerian and Lagrangian descriptions of a  closed meridian strip on a torus with the width $d$ given by  (\ref{dTore}) for $\eta^2=\delta=1\%$ (with $\vert\bK^D\vert$ in color scale). Left bottom: zoom on segment of the Lagrangian description. }
			\label{Thorus2}

\end{figure}

In Figures \ref{Thorus1} and \ref{Thorus2}, we have plotted two closed strips on a torus of radii $R_*=1, r_*=0.5$ constructed along two closed curves given in the parametric form as $t\to (\phi^0(t),\psi^0(t))$, where $s\to t(s)$ is obtained from the normalization equation  (\ref{NormTore}), and $t\in [0,2\pi]$. 

The first one, following the torus' parallels, is given by $ \phi^0(t)=\alpha t,   \psi^0(t)=t$, with $\alpha=10$.  Since the width $d$ was computed from (\ref{dTore}),  we note that the geodesic curvature $k_0^{geo}$ was much smaller than the surface curvature $\vert \K\vert$, which gives an almost uniform width. As we can see from Figure \ref{Thorus1}, the curvature deviator is very large in the central part, which means that the required pre-stress is not isotropic.   

The second one, following the torus' meridians, is given by $\phi^0(t)=t,   \psi^0(t)=\alpha t$, with $\alpha=20$ (see Figure \ref{Thorus2}).  As before, the width $d$ computed from (\ref{dTore}),  is almost uniform. The geodesic curvature $k_0^{geo}$ is rather small, which gives an overall line segment shape of the Lagragian configuration. However, the Lagrangian curve has periodic oscillations. If we zoom on one period, we note that the curvature deviator is very large, 
which means that the required pre-stress is again not isotropic. Morevoer, the obtained design shows self-intersections which from a practical point of view the technological realization are problematic. This issue, related to local minima along the relaxation path will be the subject of future research.

\subsection{Rotoidal strips}
\label{RotStrip}
Here we construct $\bU$ as a special class of ruled surface  $(s,q)\to \br(s,q)=\br_0(s)+q\bv(s)$.  The  curve $\bc \subset \R^3$  is the  directrix  and  $s \to \bv(s)  \in \R^3$ is  the direction of the generators which will be supposed to be  orthogonal on $\bc$  and described by the rotational angle $s \to \theta(s)$. As previously, $s\in (0,l)$  is the arc-length, $\btt_0(s), \bni_0(s), \bmm_0(s)$ are the tangent, the normal and the binormal unit vectors, and $k_0(s)$, $\tau_0(s)$  are the curvature and the torsion of the curve  
$\bc$.  Let $\bv(s)=\cos(\theta(s))\bmm_0(s)+\sin(\theta(s))\bni_0(s)$ and let  $\bU\subset \RR^3$ be given by its parametric description $ \bU=\{ \br(s,q) \; ; \;  s\in (0,l), q\in (-d_0(s),d_0(s)\}$, where 
$$  \br(s,q)=\br_0(s)+ q [\cos(\theta(s))\bmm_0(s)+\sin(\theta(s))\bni_0(s)],$$
$d_0$ is the width of $\bU$ and $\theta(s)$ is the angle between $\bmm_0(s)$ and the tangent plane to $\bU$ in $s.$ Obviously, the couple $(s,q)$ are orthogonal curvilinear coordinates of $\bU,$ the local basis is given by  
$\bb_s=(1-qk_0sin(\theta))\btt_0+q(\theta'-\tau_0) [\cos(\theta)\bni_0-\sin(\theta)\bmm_0], \; 
\bb_q= \cos(\theta)\bm_0+\sin(\theta)\bni_0$
and the  metric tensor is 
$L_s^2=g_{ss}(s,q)=(1-qk_0\sin(\theta))^2+q^2(\dot{\theta}-\tau_0)^2, \; g_{sq}=0, \;  L_q^2=g_{qq}=1.$
Then, the physical base is $\displaystyle \be_s=\frac{1}{L_s}\bb_s,  \be_q=\bb_q $ and the normal unit vector is given by 
$$
\be_3(s,q)=\frac{1}{L_s}\bb_s\wedge\bb_q=\frac{1}{L_s}\left(q(\dot{\theta}-\tau_0)\btt_0+(1-qk_0\sin(\theta))[-\cos(\theta)\bni_0+\sin(\theta)\bmm_0]\right)
.$$ 
The curvature tensor can now be written as 
\begin{equation} \label{curv}
\K=\frac{1}{L_s}\frac{\partial \be_3}{\partial s} \otimes \be_s+\frac{\partial \be_3}{\partial q} \otimes \be_q,  
\end{equation} 
and the geodesic curvature of the curve $\bc$ on surface $\bU$ has a simple expression: 
\begin{equation} \label{kgeo}
k_0^{geo}(s) =k_0(s)\sin(\theta(s)). 
\end{equation} 

Since $\U$ is a ruled surface,  one can get (\ref{SS})  from a direct estimation of the metric tensor (i.e.,  we do not need (\ref{SSK})). Indeed, if  we consider the  rotoid strip-surface  $$\bss_0=\{ \br(s,q) \; ; \;  s\in (0,l), q\in (-d(s),d(s)\} \subset \bU$$ along the curve $\bc$ but  with a smaller width $d(s)\leq d_0(s)$ such that  
\begin{equation}\label{SSKr}
d(s)\vert\K\vert= \Oo(\eta), \quad \quad d(s)k_0^{geo}(s)= \Oo(\eta), \quad d(s)(\dot{\theta}(s)-\tau_0(s))=\Oo(\eta), 
\end{equation} 
then  (\ref{SS})  holds.

\subsubsection{Helicoid}

Here, let us  consider  the helicoid, which is the simplest example of rotating a flat ribbon along a curve. 
For this,  let $\bc$ be a straight line in the $OX_1$ direction and define the helicoid $\bU$ by choosing $s=S=X_1, q=Q=X_2$,  and 
$\btt_0=\bc_1, \bni_0=\bc_2$ and  $\bmm_0=\bc_3$  (here $\bc_1, \bc_2, \bc_3$ the Cartesian basis in the Lagrangian configuration).  After some algebra,  we obtain 
$\bb_s = \bc_1 + q\dot{\theta}(\cos(\theta)\bc_2-\sin(\theta)\bc_3), \;
\bb_q = \sin(\theta)\bc_2+\cos(\theta)\bc_3, $ and 
$L_s^2= 1+q^2\dot{\theta}^2, \;
\be_3=\frac{1}{L_s}\left(q\dot{\theta}\bc_1-\cos(\theta)\bc_2+\sin(\theta)\bc_3\right)$.  Since $k_0=k_0^{geo}=0$,  we have $\bc=\bCC$ and 
bearing in mind that 
$$\frac{\partial \be_3}{\partial s}\cdot \bb_s= \frac{q\ddot{\theta}}{L_s}, \quad \frac{\partial \be_3}{\partial s}\cdot \bb_q= \frac{\partial \be_3}{\partial q}\cdot \bb_s=\frac{\dot{\theta}}{L_s}, \quad
\frac{\partial \be_3}{\partial q}\cdot \bb_q=0,$$
from  (\ref{curv}) we deduce that the surface-strip width conditions (\ref{SSKr}) read 
$$ d(s)\dot{\theta}(s)=\Oo(\eta),  \quad d^2(s)\ddot{\theta}(s)=\Oo(\eta).$$ 
We can  compute the Lagrangian curvature tensor $\bK$  to find 
\begin{equation} \label{K0SL}  
\bK(X_1,X_2)= \frac{X_2\ddot{\theta}(X_1)}{L_s(X_1, X_2)}\bc_1\otimes\bc_1+ 
\frac{\dot{\theta}(X_1)}{L_s(X_1, X_2)}(\bc_2\otimes\bc_1+\bc_1\otimes\bc_2).
\end{equation}

\begin{figure}[ht]
	\centering
	\includegraphics[scale=0.2]{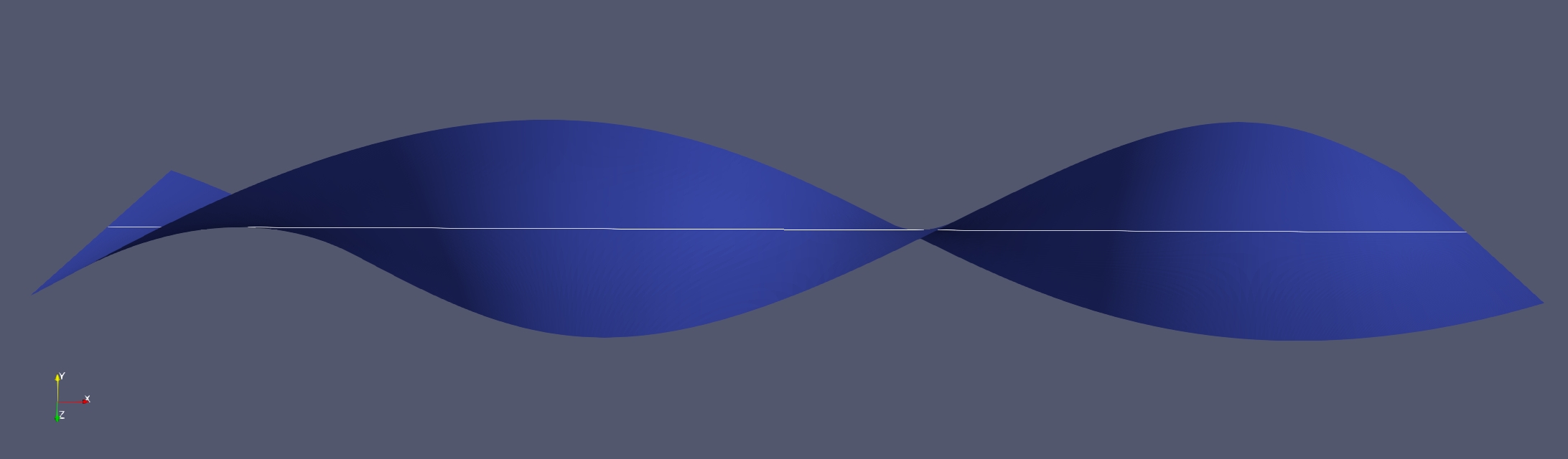}
	\includegraphics[scale=0.111]{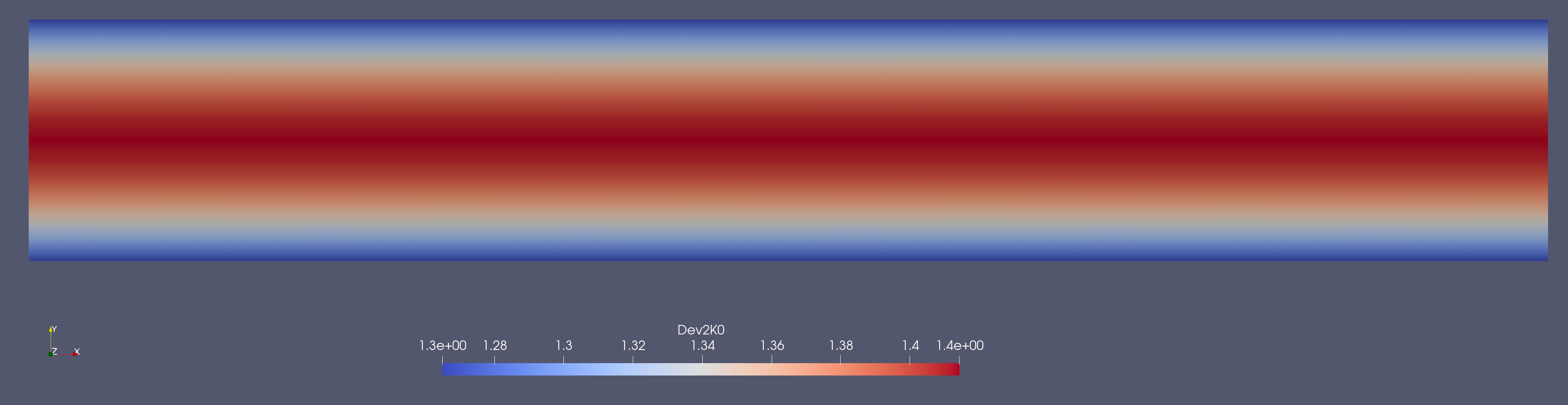}
		\caption{Top: the Eulerian description of a  helicoid strip with a constant rotating rate $\omega$. Bottom: the Lagrangian description of the strip with the  deviator norm $\vert\bK^D\vert$ in color scale.}
	\label{helicoid}
\end{figure}

As we can see in Figure \ref{helicoid},  for a constant rotating rate $\omega$, i.e.,  $\theta(S)=\omega S$,  the Lagrangian curvature tensor $\bK$ and 
the pre-stress couple given by  
\begin{equation}
\label{SMobR}
\bK(X_2)=
\frac{\omega}{\sqrt{1+\omega^2 X_2^2}}(\bc_2\otimes\bc_1+\bc_1\otimes\bc_2), \quad -\bM^*=\frac{H^3\bar{\mu}}{6}\bK(X_2).  
\end{equation}
are traceless, inhomogeneous (with respect to the width variable) and anisotropic.  
 
We notice that, depending on the rotating rate $\omega,$ one can approximate $$\bK(X_2)=
\frac{\omega}{\sqrt{1+\omega^2 X_2^2}}(\bc_2\otimes\bc_1+\bc_1\otimes\bc_2)\simeq
\omega(\bc_2\otimes\bc_1+\bc_1\otimes\bc_2)$$ recovering, in the limit of small rotation rate, the result in \cite{armon2011}. 

\subsubsection{Classical M\"{o}bius ribbon}

Another example of a non-developable surface obtained by rotating a ribbon along a curve is the classical M\"{o}bius ribbon, well-documented in the literature. 
To define it, let $\bc$ be the circle of radius $R_*$ given by 
$\br_0(s)=R_*\be_r(\phi)$ in the cylindrical 
coordinates $r,\phi, z$ with  $s=R_*\phi$ and 
$s\in (0,2\pi R_*)$. Then we have 
$\btt_0(s)=\be_\phi(\phi), 
\bni_0(s)=-\be_r(\phi), 
\bmm_0(s)=\be_z$ and $k_0(s)=1/R_*, \tau_0(s)=0$. 

A one-level M\"{o}bius ribbon $\bU$ is characterized by the choice 
$\theta(s)=\phi/2=s/(2R_*)$ but one can also consider any rotating rate $\omega$ multiple of $1/2R_*$. 
Anyway, we have $\br(s,q)=R_*\be_r(\phi)+ q [\cos(\phi/2)\be_z-\sin(\phi/2)\be_r(\phi)],$ 
with $q\in(-d_0,d_0)$, and 
$\bb_s=(1-\frac{q}{R_*}\sin(\phi/2))\be_\phi- \frac{q}{2R}
\left( \cos(\phi/2)\be_r+\sin(\phi/2)\be_z\right)$, 
$\bb_q=\cos(\phi/2)\be_z-\sin(\phi/2)\be_r,$ and we deduce $L_s^2= g_{ss}(s,q)=(1-\frac{q}{R_*}\sin(s/2R_*))^2 +\frac{q^2}{4R_*^2}$.

\begin{figure}[ht!]
	\centering
	\includegraphics[scale=0.101]{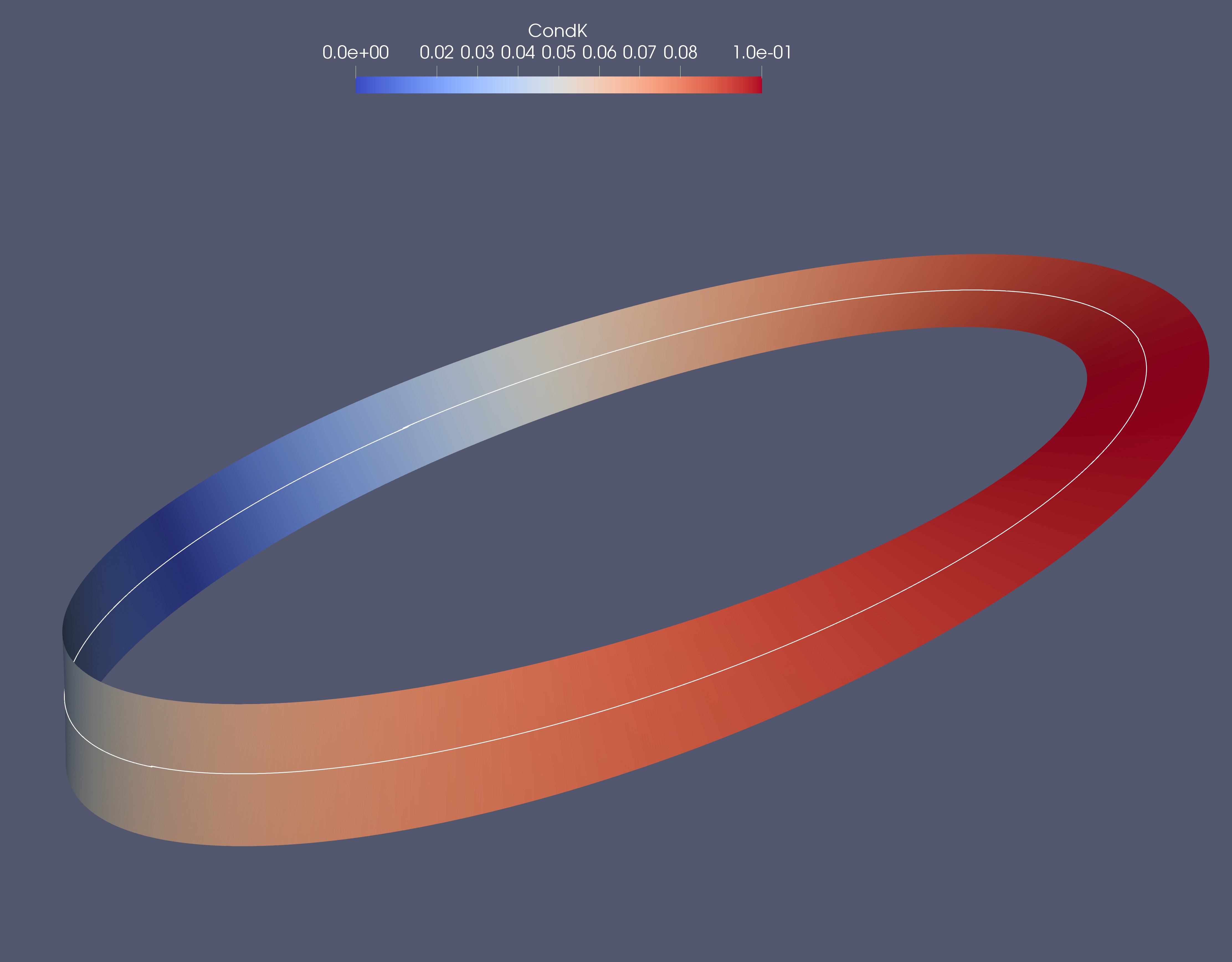}
	\includegraphics[scale=0.0501]{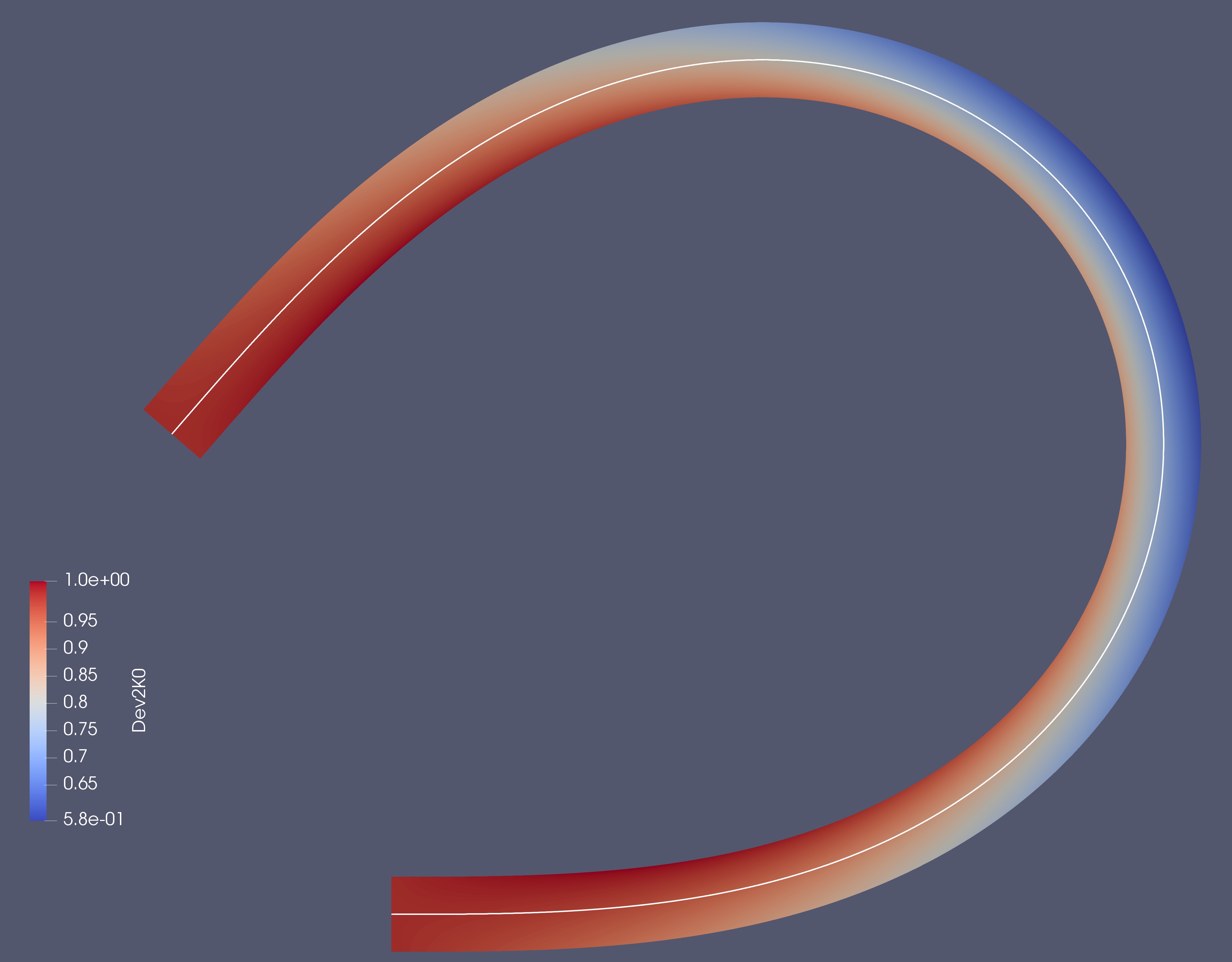}
	\includegraphics[scale=0.0501]{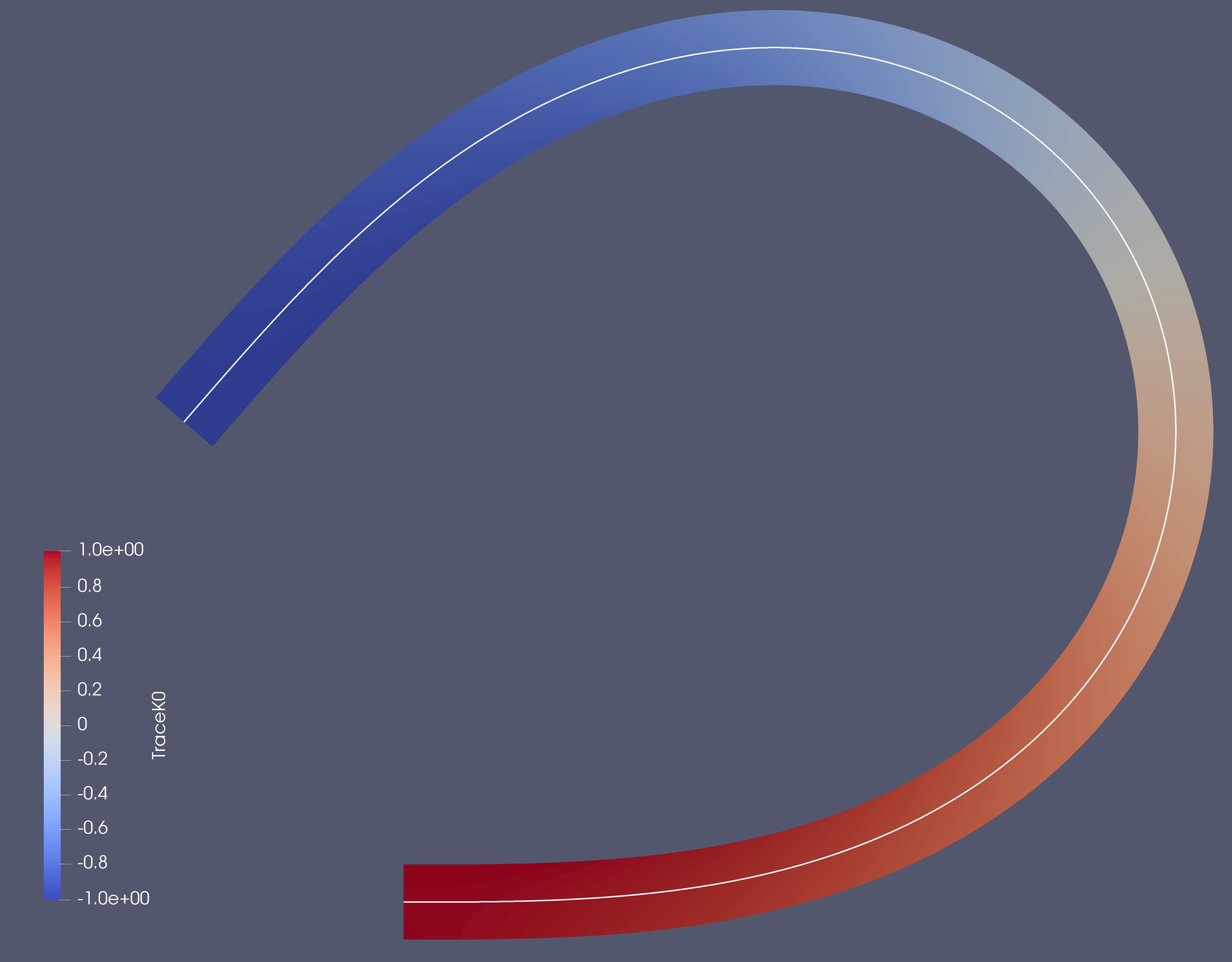}
	\caption{Top: the Eulerian description of the classical M\"{o}bius ribbon with $k_0^{geo}d=K_0d$ in color scale. Bottom: the Lagrangian description of the strip with the deviator norm $\vert\bK^D\vert$ in color scale (left) and the curvature $trace(\bK)/2$ (right). The Lagrangian and the Eulerian configurations are plotted at different length scales. 
	}
	\label{Mob_num}
\end{figure}

Since $k_0^{geo}(s)=\frac{1}{R_*}\sin(s/2R_*)$,  
the surface-strip width  conditions (\ref{SSKr}) read 
\begin{equation} \label{SSM}
\frac{d}{R_*}=\Oo(\eta),
\end{equation} 
and we can construct the Lagrangian curve $\bCC$ with the curvature $K_0(S)=k_0^{geo}(S)$. 
A straightforward computation of the Lagrangian curvature tensor gives 
$\bK=\frac{1}{R_*L_s}(1+\frac{q^2}{2R^2L_S^2})\cos(\phi/2)\be_S\otimes\be_S+ \frac{1}{2R_*L_sL_S}(\be_Q\otimes\be_S+\be_S\otimes\be_Q)$,
which can be approximated by
$$R_*\bK=\cos(\frac{S}{2R_*})(1+QK_0(S))\be_S\otimes\be_S+ \frac{1+2QK_0(S)}{2}(\be_Q\otimes\be_S+\be_S\otimes\be_Q) +\Oo(\eta^2).$$
As we can see from  Figure \ref{Mob_num}, the pre-stress associated to the designed Lagrangian strip is neither homogeneous nor isotropic. Moreover, in contrast to the developable M\"{o}bius  strip,  the shape of the Lagrangian strip is not a straight ribbon strip.

\section{Conclusions and perspectives}

The simplest relaxation experiment shows that a uniaxial pre-stressed plate relaxes toward a cylindrical shell. A natural question is then: how to extend this result to a shell of arbitrary curvature? To overcome the geometrical difficulty related to the non-developable nature of an arbitrary shell, in \cite{danescu2020shell} we replaced the isometric transformation (as is the case for cylindrical shells) with small-strains but large rotation transformations of the mid-surfaces. For many applications,  the small-strains framework is a technological restriction rather than a mathematical simplification, as we consider small pre-strains and brittle materials. The resulting design model relates the curvature of the target shell shape to the plate shape and the pre-stress distribution. 

However, even including small-strains and large rotations, isotropically pre-stressed disks (or squares) relax toward spherical caps (which are non-developable surfaces) only if their radius is small enough \cite{danescu2020shell,de2020energy}. To overcome this geometric restriction, here we consider strip-shells and design arbitrary shells by union of interconnected shell-strips. The local geometric description of an arbitrary shell involves two characteristic lengths and a natural small non-dimensional parameter $\delta\ll 1$,  which is the product between the thickness of the shell and the curvature norm. The geometries of both plate-strips and shell-strips have an additional characteristic length and thus an additional small parameter $\eta \ll 1$, which is the product between the width of the strip and the curvature of its supporting curve. We consider here a special class of strips, called the second-order strips, along an arbitrary curve on the shell mid-surface.  These strips have a specific second-order  variation with respect to the width variable so that the metric tensor can be estimated with respect to the geodesic curvature of the supporting curve. Moreover, some other additional conditions involving the width have to be fulfilled.
Then, for $\delta = \eta^2$ we prove here that the small strain condition is fulfilled, thus enabling us to obtain a simple model to design the corresponding plate-strip (i.e., to compute the shape and pre-stress moment of the plate) of any strip of a given shell.
More exactly, mapping a plate-strip to a shell-strip naturally  introduces the  difference between the  geodesic curvature of the shell  supporting curve and  the curvature of the plate supporting curve.  By designing the planar  supporting curve such that this difference is of order of $\delta$  we can control the strain  to be of order  $\delta$ on all the strip. This relation between the two curvatures  is not related to the characterization of  the asymptotic curves in differental geometry (see for instance, \cite{Spivak99}).

The resulting model was used  covers the sphere  completely and the torus partially, both of them non-developable surfaces, toward applications in photonics. Regarding the sphere as a union of meridians/parallel strips, the problem is reduced to finding the optimal number and width distribution of each strip in order to fulfill the uniform scaling conditions. For constant latitude strips, the strip width is the solution of an iterative nonlinear system for any given scale $\delta$. All these solutions present a common technological drawback, as they rely on very sharp angles (see Figures \ref{SphereMeridians} and \ref{SphereParallels}), difficult to realize by photo-lithography. 
We notice that at very low length scale ($nm$-thick multi layers),  as the available technology is only planar, the restriction to homogeneous and isotropic pre-stress is mainly a technological imperative so that an interesting open question concerns all geometries attainable by starting with the pre-stress in this class. 

Partial covering of the torus (see Figures \ref{Thorus1} and \ref{Thorus2}), the helicoid (see Figure \ref{helicoid}) and the classical M\"{o}bius ribbon  (see Figure \ref{Mob_num}) requires non-homogeneous and anisotropic pre-stres,  which is very difficult to obtain by using only the planar technology of the epitaxial growth. However, externally unloaded elastic solids can still be endogenously pre-stressed by distributed self-equilibrated force couples as shown in \cite{boley2019,kim2012,gladman2016,siefert2019} which suggests a way to produce non-isotropic and non-uniform pre-stresses, i.e. including more degrees of freedom to engineer the pre-stress. In that framework, we notice that the weak-transversal heterogeneity assumption may not be fulfilled, so that extending the presented design model to a more general setting may be necessary.

\vspace{1cm}
{\bf Acknowledgements.}  This work was partially supported by a grant 
of the French Research Agency (ANR-17-CE24-0027).

\newpage
\bibliography{Design}

%\newpage
\section{Appendix}

Here we aim to construct a surface-strip $\bss_0$ of a given surface $\bU$ along a given curve $\bc \subset \bU$ such that  $\bss_0$ t satisfies (\ref{SSK})-(\ref{SS}).  Let us consider a surface $\bU \subset \RR^3$  given by its parametric description $u \to \br_\U(u) \in \RR^3$, where $u=(u_1,u_2)$  are the parameters belonging  to $\Omega \subset \RR^2$.     We denote by  $\bb_1=\partial_{u_1} \br_\U,  \bb_2=\partial_{u_2} \br_\U$  the covariant basic vectors  and by   $g_{11}=|\bb_1|^2, g_{22}=|\bb_2|^2, g_{12}=\bb_1\cdot \bb_2$ the covariant fundamental magnitudes of the first order.   Let $\T=\T(u)=Sp\{\bb_1, \bb_2\}$  be the two-dimensional vector space tangent to the surface $\bU$. We also denote by $g=\sqrt{g_{11}g_{22}-g_{12}^2}$ the element of area  and by   $\be_3=\bb_1 \wedge \bb_2/g $ the unit normal of $\bU$.  We introduce the contravariant tangent basis, denoted by  $\bb^1,\bb^2$, and the contravariant fundamental magnitudes of the first order $g^{11}=|\bb^1|^2=g_{22}/g^2, \quad g^{22}=|\bb^2|^2=g_{11}/g^2, \quad g^{12}=\bb^1\cdot \bb^2=-g_{12}/g^2$.   

On this surface, we consider a non-planar curve $\bc \subset \bU $ given by its parametric description $s \to \br_0(s)=\br_\U(u^0(s))  \in \bU$, where $s\in (0,l)$ is the arc-length.   Let $\btt_0(s)=\frac{d}{d s} \br_0(s)=\nabla\br_\U \frac{d}{d s}u^0(s)= \frac{d}{d s}u_i^0(s)\bb_i(u^0(s))  \in \T(u^0)$ be the tangent unit vector on the curve.  
%We denote by $\bn_0(s)$ the unit normal on  $\btt_0$, i.e.,  $$\frac{d}{d s}\btt_0(s)=k_0(s)\bn_0(s), \quad  \vert \bn_0(s) \vert =1,  \quad \bn_0(s) \cdot  \btt_0(s)=0$$ where $k_0(s)$ is the curvature of  $\bc$, and by $\bm_0(s)= \btt_0(s)\land \bn_0(s)$  the binormal vector. 
We denote by $ \bni_0^\U(s)=\be_3(u^0(s))\wedge  \btt_0(s)$ the unit vector which has the direction of the projection of $\bni_0(s) $ in the plane $ \T(u^0)$, i.e.,  $ \bni_0^\U(s)$ belongs to the intersection of the normal plane on $\bc$ with the tangent plane of the surface  $\bU$, and   $\vert \bni_0^\U(s)\vert =1,   \bni_0^\U(s)\cdot \btt_0(s)=0 $.  From the above Frenet formula we get 
\begin{equation} \label{nu0}
	\frac{d \bni_0^\U(s)}{d s}\cdot \btt_0(s)=
	-k_0(s)(\bni_0(s) \wedge  \be_3(u^0(s)))\cdot 
	\btt_0(s)=-k_0(s)\be_3(u^0(s))\cdot \bmm_0(s)
\end{equation}
and we recognize here {\em the geodesic curvature} $k_0^{geo}$  of the curve $\bc \subset \bU$, given by (\ref{kgeo}).

Let $s \to v^0(s)$ be such that 
\begin{equation} \label{v0}
	\bni_0^\U(s)=\nabla\br_\U v^0(s) = v_i^0(s)\bb_i(u^0(s)).  
\end{equation}
Then we can define the strip $\bss_0 \subset \bU$ given by (\ref{sss})
where $2d(s)$ is the strip "width" and $w^0$ will be determined later.   Denoting by 
$u_i(s,q)=u^0_i(s)+qv^0_i(s)+\frac{q^2}{2}w^0_i(s)$, by $ \dot{u}_i(s,q)=\partial_su_i(s,q)=\dot{u}^0_i(s)+q\dot{v}^0_i(s)+\frac{q^2}{2}\dot{w}^0_i(s)$ and by  $u_i'(s,q)=\partial_qu_i(s,q)=v^0_i(s)+qw^0_i(s)$,   the local basis is given by  
$$\bb_s(s,q)=\frac{\partial}{\partial  s}\br(s,q)=\dot{u}_i(s,q) \bb_i(u(s,q)), \quad\bb_q(s,q)=\frac{\partial}{\partial  q}\br(s,q)=u_i'(s,q) \bb_i(u(s,q)),$$
while the metric tensor is given by 
%%%%%%%%%%%%%%%%%%
\begin{equation} \label{gss0}
	g_{ss}(s,q)=g_{ij}(u(s,q))\dot{u}_i(s,q) \dot{u}_j(s,q), \quad
	g_{sq}=g_{ij}(u(s,q))\dot{u}_i(s,q) u_j'(s,q), 
\end{equation} 
\begin{equation} \label{gqq0} g_{qq}=g_{ij}(u(s,q)) u_i'(s,q)u_j'(s,q).
\end{equation} 
If we suppose now that 
\begin{equation}\label{gas}
	d(s)\frac{\partial g_{ij}}{\partial u_k}(u^0(s))=  \Oo(\eta), \quad d^2(s)\frac{\partial^2 g_{ij}}{\partial u_k\partial u_l}(u^0(s))=  \Oo(\eta^2),
\end{equation}
then we get 
$$g_{ij}(u(s,q)) =g_{ij}(u^0(s)) +q\frac{\partial g_{ij}}{\partial u_k}(u^0(s))v^0_k(s) +  \Oo(\eta^2).$$
This last estimation and the following assumptions
$$d(s)\frac{\partial g_{ij}}{\partial u_k}(u^0(s))v^0_k(s)\frac{d u_i^0}{d s} (s)\frac{d u_j^0}{d s} (s)= \Oo(\eta), \quad \mbox{for all} \; s\in(0,s),$$
$$d(s)\frac{\partial g_{ij}}{\partial u_k}(u^0(s))v^0_k(s)\frac{d u_i^0}{d s}(s) v_j^0(s) = \Oo(\eta), \quad \mbox{for all} \; s\in(0,s),$$
$$d(s)\frac{\partial g_{ij}}{\partial u_k}(u^0(s))v^0_k(s)v^0_i(s)v^0_j (s)= \Oo(\eta), \quad \mbox{for all} \; s\in(0,s),$$
can be used  to estimate the  metric tensor from (\ref{gss0}-\ref{gqq0}) 
$$g_{ss}=g_{ij}(u^0(s))\frac{d u_i^0}{d s}\frac{d u_j^0}{d s}+ q\left( 2g_{ij}(u^0(s)) \frac{d u_i^0}{d s}\frac{d v_i^0}{d s}  + \frac{\partial g_{ij}}{\partial u_k}(u^0(s))v^0_k \frac{d u_i^0}{d s} \frac{d u_j^0}{d s} \right)+  \Oo(\eta^2)$$
$$ g_{sq}=g_{ij}(u^0(s)) \frac{d u_i^0}{d s} v_j^0 +q \left( g_{ij}(u^0(s))\frac{d v_i^0}{d s}v_j^0+ g_{ij}(u^0(s))\frac{d u_i^0}{d s}w_j^0 +  \frac{\partial g_{ij}}{\partial u_k}(u^0(s)) v^0_k  \frac{d u_i^0}{d s} v_j^0 \right)+  \Oo(\eta^2), $$ 
$$g_{qq}=g_{ij}(u^0(s)) v_i^0 v_j^0 +q\left( 2g_{ij}(u^0(s))w^0_i v_j^0+  \frac{\partial g_{ij}}{\partial u_k}(u^0(s)) v^0_k v^0_iv^0_j  \right)+  \Oo(\eta^2).$$
We now choose $w^0$ such that the first order term (i.e., which multiplies $q$) vanishes in expression of $g_{qq}$, i.e., we put $w_0$ given by (\ref{w0}). Then, bearing in mind that $g_{ij} v_i^0 v_j^0=\vert \bni_0^\U(s)\vert^2 =1$
we get 
\begin{equation}\label{gqq}
	g_{qq}=1+  \Oo(\eta^2).
\end{equation}
Using (\ref{w0}) again, we get 
$$ g_{ij}\frac{d v_i^0}{d s}v_j^0+ g_{ij}\frac{d u_i^0}{d s}w_j^0 +  \frac{\partial g_{ij}}{\partial u_k}v^0_k  \frac{d u_i^0}{d s} v_j^0= g_{ij}\frac{d v_i^0}{d s}v_j^0+\frac{1}{2}\frac{\partial g_{ij}}{\partial u_k}v^0_k  \frac{d u_i^0}{d s} v_j^0= \frac{d   \bni_0^\U }{d s}\cdot  \bni_0^\U=0, $$ 
hence the first order term vanishes in expression of $g_{sq}$ and  since $\displaystyle g_{ij} \frac{d u_i^0}{d s} v_j^0 =   \bni_0^\U(s)\cdot \btt_0(s)=0$ we get 
\begin{equation}\label{gsq}
	g_{sq}=\Oo(\eta^2).
\end{equation}
If we note that 
$$g_{ij}\frac{d u_i^0}{d s}\frac{d u_j^0}{d s}=\vert \btt_0 \vert^2=1, \quad  2g_{ij} \frac{d u_i^0}{d s}\frac{d v_i^0}{d s}  + \frac{\partial g_{ij}}{\partial u_k}v^0_k \frac{d u_i^0}{d s} \frac{d u_j^0}{d s}=2\frac{d   \bni_0^\U }{d s}\cdot  \btt_0, $$ 
we obtain
\begin{equation}\label{gss} g_{ss}(s,q)=1-2qk_0(s)\be_3(u^0(s))\cdot \bmm_0(s)+ \Oo(\eta^2).\end{equation}
From (\ref{gqq}-\ref{gss}) we conclude that if  (\ref{SSK})  holds, then   subsurface $\bss_0$ of $\bU$, given by (\ref{sss}),  is a strip-surface, i.e.,  (\ref{SS}) holds. 

\end{document}